\documentclass[fleqn,usenatbib]{mnras}

\usepackage{newtxtext,newtxmath}

\usepackage[T1]{fontenc}

\DeclareRobustCommand{\VAN}[3]{#2}
\let\VANthebibliography\thebibliography
\def\thebibliography{\DeclareRobustCommand{\VAN}[3]{##3}\VANthebibliography}

\usepackage{graphicx}	
\usepackage{amsmath}	
\usepackage{pdflscape}

\usepackage[usenames,dvipsnames]{xcolor}

\newcommand\dDLR{d_\text{DLR}}
\newcommand\Nser{n_\text{Ser}}

\title[Hostless supernovae]{The Statistics and Environments of Hostless Supernovae}

\author[Qin et al.]{
Yu-Jing Qin,$^{1}$\thanks{E-mail: yujingq@caltech.edu}
Ann Zabludoff,$^{2}$
Iair Arcavi$^{3}$
Nathan Smith$^{2}$
Yakov Faerman$^{4}$
and Dan Maoz$^{5}$
\\
$^{1}$California Institute of Technology, 1200 E California Blvd, Caltech MC 249-17, Pasadena, CA 91125\\
$^{2}$Department of Astronomy and Steward Observatory, University of Arizona, 933 N Cherry Ave, Tucson, AZ 85721\\
$^{3}$The School of Physics and Astronomy, Tel Aviv University, Tel Aviv 69978, Israel\\
$^{4}$Department of Astronomy, University of Washington, Seattle, WA 98195, USA\\
$^{5}$The School of Physics and Astronomy, Tel Aviv University, Tel Aviv 69978, Israel
}

\date{Accepted XXX. Received YYY; in original form ZZZ}

\pubyear{2024}

\begin{document}
\label{firstpage}
\pagerange{\pageref{firstpage}--\pageref{lastpage}}
\maketitle

\begin{abstract}
Transient surveys routinely detect supernovae (SNe) without obvious host galaxies.
To understand the demographics of these `hostless' SNe and to constrain the possible host properties, we identify $161$ SNe reported to the Transient Name Server since 2016 that do not have hosts cataloged from pre-explosion wide-field galaxy surveys. 
Using forced aperture photometry, we detect excess flux around only 56 of these SNe.
Both thermonuclear and core-collapse (CC) SNe are present in our sample.
Compared to flux-limited SNe samples with known hosts, superluminous supernovae (SLSNe), particularly hydrogen-deficient SLSNe, are over-represented here relative to all other SNe types; among CC SNe, there is also a higher fraction of interacting SNe than non-interacting.
On the low-luminosity side, seven SNe have host absolute magnitude upper limits fainter than $M_g=-12$, about $1$ per cent of the Small Magellanic Cloud's luminosity; the faintest limits are close to the luminosity of globular clusters or ultra-faint dwarf galaxies ($M_g\simeq -8$).
Fitting multi-band forced photometry, 11 SNe have host stellar masses $<10^6$ M$_{\odot}$ assuming quiescent hosts, and 13 SNe have host stellar masses $<10^5$ M$_{\odot}$ assuming star-forming hosts.
The spatial distribution of hostless SNe indicates that the majority are not associated with known galaxy groups and clusters, ruling out intracluster stellar light as the primary contributor of such SNe. Hostless Type Ia SNe tend to be more luminous and slow-fading than SNe Ia with known host galaxies, implying a hidden population of low-mass and star-forming hosts. We conclude that any undetected host galaxies are likely star-forming dwarfs in the field.
\end{abstract} 

\begin{keywords}
transients: supernovae -- supernovae: general -- galaxies: statistics
\end{keywords}



\section{Introduction} \label{sec:intro}

Supernovae (SNe) are cataclysmic explosions of stars and are thus commonly observed in galaxies. However, untargeted transient surveys routinely find SNe without apparent host galaxies in the discovery or archival survey images.
These `hostless' or `orphan' SNe account for a non-negligible fraction of SN detections. For example, the flux-limited Zwicky Transient Facility Bright Transient Survey (ZTF BTS; \citealt{Fremling20}) found a hostless fraction of $2.2\pm0.6$ per cent for Type Ia supernovae (SNe Ia) and a comparable fraction of $1.5^{+1.1}_{-0.7}$ per cent for core-collapse supernovae (CC SNe) \citep{Perley20}. The Dark Energy Survey Supernova Program reported a similar hostless fraction of $\sim 2$ per cent \citep{Smith20des}.
Earlier results show even higher fractions of hostless SNe, e.g., $\sim 4$ per cent in the Sloan Digital Sky Survey-II Supernova Survey \citep{Sako18} and nearly $7$ per cent in the Supernova Legacy Survey \citep{Sullivan06}.
While these hostless fractions depend on the criteria of host association and the sensitivity limits of galaxy and SN surveys, hostless SNe represent a relatively rare yet notable family of transients.

Despite the growing number of cases contributed by ongoing transient surveys, a general demographic study of hostless SNe, including their cosmic environments and possible host properties, is still missing.
One reason is that transient surveys do not routinely report host galaxies, or the absence thereof, for their SN detections, and it is non-trivial to determine which SNe are hostless.
Also, deep \textit{and} wide-area imaging surveys, which are required to detect fainter hosts or put rigorous limits on host properties, have only become available in recent years.
With more discoveries imminent, particularly by the Legacy Survey of Space and Time (LSST; \citealt{Ivezic19}) at Vera C. Rubin Observatory, a pilot study of hostless SNe from archival data is timely and necessary to further develop our picture of transient demographics.

Hostless SNe may arise for various reasons.
The most likely situation is that their hosts are too faint to be detected.
State-of-the-art wide-area sky surveys have extended source limiting magnitudes ($m_\text{lim}$) of $\sim23$ to $24$ mag in optical bands (e.g., \citealt{Dey19, SevillaNoarbe21}).
At such sensitivity, a faint dwarf galaxy of absolute magnitude $M\sim-12$ (typical of dwarf satellite galaxies in the Local Group; e.g., \citealt{McConnachie12}) will become undetectable at a redshift of merely $z=0.023$ ($m_\text{lim}\sim23$) or $0.036$ ($m_\text{lim}\sim24$), assuming no extinction.
However, the SNe luminosity function (e.g., \citealt{Li11lumfunc}; \citealt{Perley20}) substantially overlaps that of galaxies; the most luminous SNe can even reach peak absolute magnitudes of $M_g\sim-21$ to $-22$ mag \citep{GalYam19}. Therefore, many SNe can outshine their host galaxies and become detectable in transient surveys, leaving their faint, diffuse hosts beneath the noise floors of galaxy surveys.
Indeed, deeper imaging data may reveal the previously unseen host galaxies of some SNe (e.g., \citealt{Zinn12, Nucita17}) and reduce the hostless fraction of a SN survey \citep{Smith20des}, particularly for high-redshift SNe.

Another situation that may lead to `hostless' SNe is that SNe can explode at extreme separations from the centre of their host galaxies.
For example, SNe can occur at the tip of a tidal tail arising from a galaxy-galaxy interaction (e.g., \citealt{Ferretti17}); SNe associated with escaped hypervelocity stars may also lead to extreme SN-host distances \citep{Zinn11hvs}.
Some transient types, particularly Calcium-rich gap transients, prefer such extreme transient-host distances (e.g., \citealt{Kasliwal12, Lyman16}), indicating that they are likely kicked out during dynamical interactions \citep{Lyman14} or ejected from galactic nuclei \citep{Foley15}.
The host galaxies of such SNe may eventually become unrecognizable when separations are large enough, contributing to the population of hostless SNe.

Hostless SNe may also originate from the population of intergalactic stars.
Tidal interactions of galaxies can strip stars away; over cosmic time, such stars build up the diffused stellar component in clusters or groups of galaxies, i.e., the intracluster light (ICL; e.g., \citealt{Gregg98, Gonzalez05, Mihos05, Gonzalez13, DeMaio18}).
Depending on the criteria for ICL, cluster mass, and redshift, the fraction of ICL contribution to the total luminosity (or stellar mass) ranges from a few per cent to over half. (e.g., \citealt{Lin04, Gonzalez05, Krick07, Seigar07, Burke15, Mihos17, JimenezTeja18, Montes18}), favouring a high fraction of ICL-associated, hostless SNe in cluster environments. 
Indeed, cluster-targeted SN surveys have detected SNe associated with the ICL (e.g., \citealt{Sand11, Graham15}). Even the mass fraction of intracluster light can be constrained using hostless SNe \citep{McGee10}.

In any of these scenarios, hostless SNe (and hostless transients in general) are effective tracers of stellar mass and star formation beyond the flux or surface brightness limits of existing wide-field galaxy surveys \citep{Conroy15}.
Knowing their demographics and environments is thus a key step in characterizing the biasing factors of such tracers, including the preference for star-forming or quiescent hosts, cluster or field environments, and galaxies or intergalactic stars.
For example, association with cluster environment would suggest an ICL contribution to this population; 
CC SNe in cluster environments may be evidence of in-situ star formation in the ICL \citep{Puchwein10}; whereas finding \textit{only} SNe Ia would suggest faint quiescent hosts.

Hostless SNe remain an important piece of the puzzle towards a comprehensive picture of the dynamic sky.
In this paper, we identify a sample of $161$ hostless SNe, constrain their host properties, and characterize their environments.
In Section \ref{sec:sample}, we describe the parent transient sample and galaxy catalogues, as well as our technique and criteria to identify hostless SNe. 
In Section \ref{sec:samplestat}, we discuss the subtype distribution, or demographics, of hostless SNe. 
In Section \ref{sec:limits}, we use the sensitivity limits of the galaxy survey images, multi-band forced aperture photometry, and comparisons of the fluxes to stellar population synthesis models to constrain the properties of the unseen host galaxies, if they exist.
Section \ref{sec:cluster} characterizes the cosmic environments of the hostless SNe and tests their association with groups or clusters.
Section \ref{sec:lc} presents the light curve properties of hostless Type Ia SNe and the implications for the host properties. 
We summarize our results in Section \ref{sec:summ}.
Through the paper, we assume a flat $\Lambda$-CDM cosmology with $H_0=70\,\mathrm{km\,s^{-1}\,Mpc^{-1}}$, $\Omega_\text{M}=0.3$.
We use the AB magnitude system throughout the paper.

\section{Sample Selection} \label{sec:sample}

\begin{figure*}
\centering
\includegraphics[width=\linewidth]{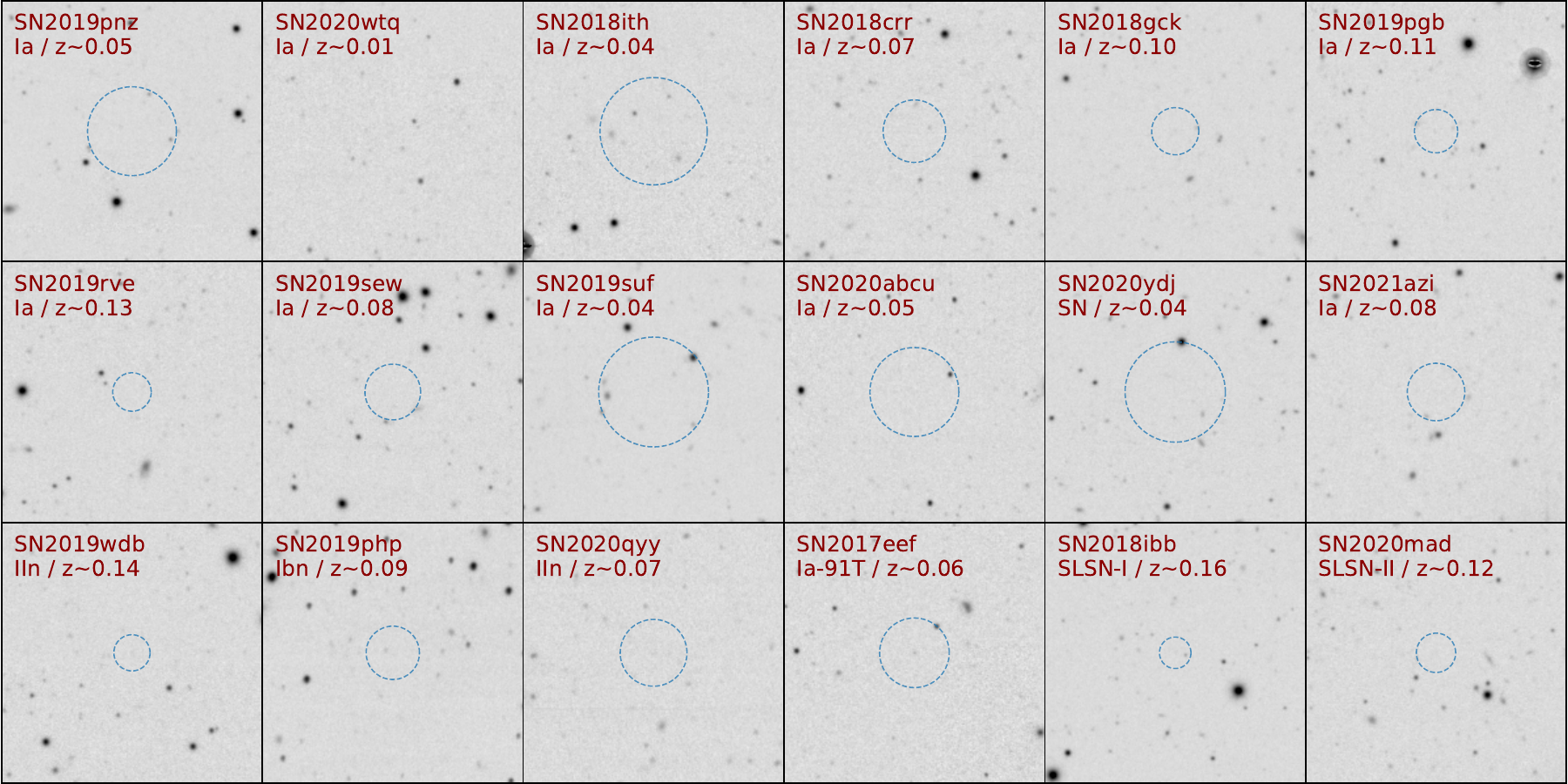}
\caption{
Examples of hostless SNe in our sample. Each panel shows the stacked $g$-band image from the DESI Legacy Surveys \citep{Dey19}.
These $2$ arcmin-sized cutouts are centred on the SN coordinates, with circles showing the $10$ kpc projected radius at the SN redshifts. 
The names, types, and redshifts of the SNe are indicated in red text.
We show examples of both thermonuclear and CC SNe to illustrate the diverse population of hostless transients.
Near the SN coordinates, there are no obvious host galaxies; other sources are either stars or at a projected distance from the SN that is large relative to the source size, making them unlikely to be the true hosts. 
\label{fig:fieldexamples}}
\end{figure*}

\subsection{Data sources and workflow}

We accessed public transient records at the Transient Name Server (TNS)\footnote{\url{https://www.wis-tns.org/}} discovered from 2016 Jan 1 to 2021 Jun 15, which include $8137$ SNe with spectroscopic classification.
This work aims to study SNe \textit{without} host galaxies; therefore, we exclude $2430$ SNe with hosts already identified and reported by transient surveys or follow-up programs.
The remaining $5707$ SNe form our parent sample.

Many transient surveys and follow-up programs do not routinely identify and report host galaxies in their workflow. 
To further exclude SNe with possible unreported host galaxies, we first attempt to identify hosts for the remaining $5707$ SNe using source catalogues of wide-field surveys (Section \ref{sec:criteria}).
For SNe without host galaxies identified in source catalogues, we visually inspect postage stamps from sky surveys to confirm the absence of their hosts (Section \ref{sec:inspection}).

We choose the DESI Legacy Imaging Surveys DR9 (LS; \citealt{Dey19}) and Pan-STARRS (PS1; \citealt{Chambers16}) $3\pi$ Steradian Survey DR2 \citep{Flewelling20} catalogs to identify possible unreported hosts.
The LS source catalogue combines data acquired from three surveys: the Mayall $z$-band Legacy Survey (MzLS), Beijing-Arizona Sky Survey (BASS; \citealt{Zou17}), and Dark Energy Camera Legacy Survey (DECaLS), which also includes images from the Dark Energy Survey (DES; \citealt{DES16}).
These surveys cover most of the high Galactic latitude region of both hemispheres in the \textit{grz} bands. In contrast, PS1 covers the sky northward to $\text{Dec}\simeq-30\degr$, including low Galactic latitude regions avoided by DESI LS, in the \textit{grizy} bands.

We search SN coordinates in the LS \textsc{tractor} \citep{Lang16} and PS1 stacked object catalogues with a fixed search radius of $1$ arcmin.
Given the better sensitivity of LS in the \textit{grz} bands, we choose the LS \textsc{tractor} catalogue to identify possible unreported hosts where complete three-band coverage is available. Outside the coverage of the LS, we use the PS1 stacked object catalogue to identify hosts.
Due to overlap in the time frame, some SNe in our parent sample also appear in the LS images and are even catalogued as point sources, hindering the detection (or exclusion) of fainter host galaxies.
PS1 images, on the other hand, were mostly acquired prior to early 2016 and are thus uncontaminated by SNe in our parent sample, enabling the detection of faint hosts that may be outshined in contaminated LS images.
We flag cases in which the SN coordinate matches a LS point source within $2$ arcsec \textit{or} cases in which the SNe themselves are visible in the LS images. These SNe will be visually inspected later using pre-explosion PS1 images.

Host galaxies are extended sources in nature.
We use the pipeline profile-fitting morphological classification in the LS catalogue and the machine learning classification of \citet{Tachibana18} (with the recommended \texttt{ps\_score} threshold of $0.829$) for the PS1 catalogue to select extended sources.
As usual, distinguishing galaxies from stars is challenging at the survey limits. The cores of compact galaxies might be identified as point sources in seeing-limited images; stars may also be confused with extended sources at low signal-to-noise ratios.
We identify unreported hosts in extended sources but also consider the possible confusion of stars and compact galaxy cores later during our visual inspection.

The LS catalogue also has an ancillary photometric redshift catalogue available \citep{Zhou21}.
The uncertainty in the photometric redshift is relatively large compared to that of the SN spectroscopic redshift, making it a less rigorous criterion to identify or exclude host galaxies than position-based matching.
Therefore, we perform two-dimensional position-based matching first and only use photometric redshifts during our later visual inspection to resolve uncertain cases.
Such a position-based method may erroneously associate genuine hostless SNe with foreground or background galaxies; as a result, our sample of hostless SNe is more pure than complete.

\subsection{Host association criteria} \label{sec:criteria}

To identify and exclude SNe with possible unreported host galaxies in the TNS, we check whether any extended source in the vicinity could be the host.
Previous works associate SNe with candidate host galaxies using the Directional Light Radius (DLR) method (e.g., \citealt{Sullivan06, Gupta16}). The angular distance of the SN to a nearby galaxy is normalized by the galaxy angular size (usually the effective radius, $r_\text{e}$) along the direction to the SN.
This normalized distance ($\dDLR$) measures the SN-galaxy separation in units of galaxy size, taking projection effects or galaxy ellipticity into consideration.
Generally, a cut-off threshold for $\dDLR$ is chosen to optimize the fraction of properly identified hosts for a SN sample, while not causing too many mismatches when the true host is undetected.
Only galaxies within this $\dDLR$ threshold can be considered as the host, and the galaxy with the minimal $\dDLR$ is then selected as the host.
When no galaxy has a $\dDLR$ lower than the threshold, the SN is considered hostless at the sensitivity limit of the galaxy survey.

The DLR method, although widely adopted for its simplicity, may suffer from biases related to galaxy light profiles.
Galaxy light profiles span a wide range in concentration, commonly parameterized using the S\'ersic index ($\Nser$; \citealt{Sersic63}).
For more concentrated light profiles (higher $\Nser$), the surface brightness drops faster in the inner region, but slower in the outskirts.
As a result, the same $\dDLR$ threshold encloses different fractions of stellar light and hence SN rates, depending on the light profile.
For example, when using $r_e$ to calculate $\dDLR$, a constant threshold of $\dDLR=4$ encloses $99.1$ per cent of the stellar light for an exponential profile ($n_{\text{Ser}}=1$), but only $84.7$ per cent of the light in the more concentrated de Vaucouleurs' profile ($\Nser=4$), making the threshold more severe in excluding hosts with exponential light profiles.
Therefore, the $\dDLR$ threshold must be adjusted for the light profile. We describe a new, adaptive-threshold DLR method based on the light profile concentration in Appendix \ref{appendix:threshold}.

For SNe with three-band coverage in the LS, we calculate $\dDLR$, following the definition in \citet{Sullivan06} and \citet{Gupta16}, for all extended sources using their effective radius.
For SNe in PS1, we calculate a $\dDLR$-like parameter for all extended sources using the $r$-band Kron radius \citep{Kron80} and the second moment-based source ellipticity.
Note that the Kron radius is defined using the first radial moment of light, instead of the enclosed fraction of light as for the effective radius. Moreover, the measured Kron radius and the moments of light in the PS1 catalogue contain the contribution of the point spread function (atmospheric and instrumental).
To ensure comparable selection criteria across the PS1 and LS catalogues, we train k-Nearest Neighbours (kNN) regressors to derive the threshold for the Kron radius-based $\dDLR$, using galaxies simultaneously imaged by the LS and PS1 as the training dataset. We describe the details in Appendix \ref{appendix:threshold}; the sample of unreported hosts we identify is inspected later to ensure the accuracy of our method.

In summary, we select a short list of $221$ candidate hostless SNe using our adaptive-threshold DLR method, in which $123$ candidates have both LS and PS1 coverage, $96$ candidates have only PS1 coverage, and two candidates have only LS coverage.

\subsection{Visual inspection} \label{sec:inspection}

The adaptive-threshold DLR method described in the previous subsection may still miss some unreported host galaxies. 
Some host galaxies lie beyond the $1$ arcmin search radius. Rarely, the confusion in star/galaxy separation or errors in source shape and size measurement can also hinder host galaxy identification.
Therefore, we perform a comprehensive visual inspection of the hostless SN candidates selected earlier to ensure that no potential host has been missed for any reason.

We inspect LS and PS1 postage stamps of $2$ and $8$ arcmin widths centred on the $221$ hostless SN candidates found with our adaptive-threshold DLR method.
We access \textit{grz}-band colour composite images from the LS using their Sky Browser image cutout service\footnote{\url{https://www.legacysurvey.org/viewer}}; we also access $r$-band stacked \textsc{FITS} images from PS1, which balances sensitivity and resolution, using their official image cutout service\footnote{\url{http://ps1images.stsci.edu/cgi-bin/ps1cutouts}}.
In the $2$ arcmin fields, we mark point and extended sources in the corresponding source catalogue (LS or PS1), including angular sizes, magnitudes, and $\dDLR$ parameters. The annotated images are juxtaposed with the original ones to facilitate inspection.
Primarily, we use the LS images for their better sensitivity over PS1 images. When there is a coincident LS point source at the SN position, the SN itself is visible in the LS images, or the LS images are affected by data coverage or quality issues (detector gaps, field edges, artefacts, etc.), we use the PS1 images for inspection.
There are $40$ candidates inside the LS footprint that are inspected using the PS1 images instead for such reasons.
For SNe outside the sky coverage of the LS, we use PS1 images for inspection.

Our inspection focuses on 1) if there are unreported host galaxies within the $1$ arcmin search radius that are \textit{not} properly identified during our initial selection using the adaptive-threshold DLR method (Section \ref{sec:criteria}; Appendix \ref{appendix:threshold}), and 2) if there are possible host galaxies beyond the $1$ arcmin search radius used for our initial selection.
We categorize hostless SN candidates into four cases:
\begin{enumerate}
    \item[-] `Good' (126 SNe), if there is no apparent host per our visual inspection, even in larger postage stamps of $8$ arcmin width or in the LS Sky Browser;
    \item[-] `Likely' (35 SNe), if there is no host per our adaptive-threshold DLR threshold, but visual inspection reveals a marginal host candidate inside the search radius;
    \item[-] `Distant' (18 SNe), if a possible host lies outside the search radius and has a photometric redshift $95$ per cent confidence interval that includes the SN redshift
    or a spectroscopic redshift consistent with that of the SN (difference $<0.01$);
    \item[-] `Failed' (42 SNe), if the host is visible in the $1$ arcmin search radius, but was not identified for any reason by the adaptive-threshold DLR method during our initial selection.
\end{enumerate}

We use galaxy photometric or spectroscopic redshifts in the LS Sky Browser to resolve cases in which possible hosts lie beyond the search radius and are thus missed by our adaptive-threshold DLR method.
To exclude low-redshift candidates with potentially large SN-host angular distances, the inspection using spectroscopic redshifts is not limited to galaxies in the $8$ arcmin width postage stamps.
Among 12 possible hosts outside the search radius, nine are identified by the coincidence of spectroscopic redshift, and three are identified by the consistency of photometric redshift.

There are cases in which the adaptive-threshold DLR method cannot identify likely hosts in the search radius.
For example, the survey may fail to detect and catalog the host, the host could be erroneously classified as a star, and the shape and size parameters might be absent.
Whenever a clear host in the search radius is visible, we consider the case a `Failed' one and exclude the SN from our sample.

Furthermore, to assess the accuracy of our adaptive-threshold DLR method in identifying unreported hosts, we also inspect a sample of $100$ SNe with hosts identified during our candidate selection (Section \ref{sec:criteria}), using similar cutouts of $2$ and $8$ arcmin widths.
Half of these unreported hosts are identified using the LS catalog, and the other half are identified using the PS1 catalog and our trained kNN regressors.
All these identified hosts are apparently reliable.

\subsection{Sample summary} \label{sec:samplesummary}

Based on visual inspection of postage stamps, we assemble two samples from candidates selected using the adaptive-threshold DLR method: the Primary Sample (126 SNe) including only `Good' cases and an Extended Sample (161 SNe) including both `Good' and `Likely' cases. The resulting sample of hostless SNe is tabulated in Appendix \ref{appendix:sources}.
To verify our sample selection procedure, we applied forced photometry within a $5$-arcsec diameter aperture at the SN position, using the same pre-explosion LS and PS1 images that we inspected earlier (Section \ref{sec:inspection}); only 56 SNe have surrounding $3\sigma$ flux excesses in any band, indicating possible marginal-level host detections.
The analysis in this paper is mainly based on the Extended Sample, in which all SNe satisfy our $\dDLR$ threshold for being hostless. A chance remains that some SNe have hosts at distances well beyond the $\dDLR$ threshold that we choose in Appendix \ref{appendix:threshold}.

\section{Results and Discussion}

\subsection{Hostless demographics} \label{sec:samplestat}

\begin{figure}
\centering
\includegraphics[width=\linewidth]{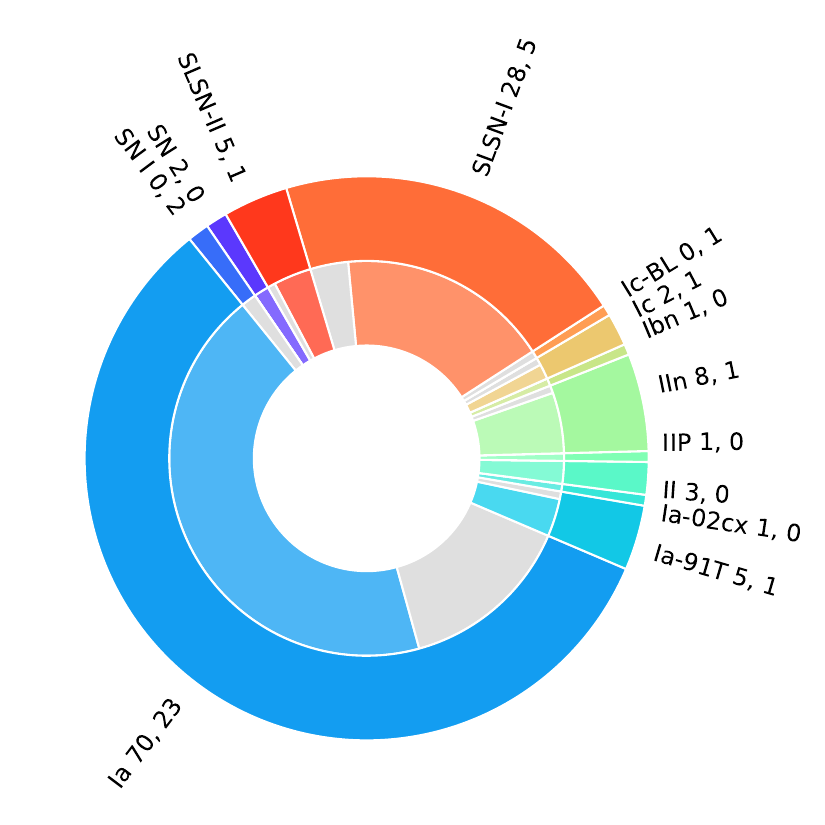}
\caption{
Number of hostless SNe under each subtype.
We show our Extended Sample in the outer ring with labels indicating the numbers of `Good,'  `Likely' cases, respectively.
The inner ring shows the breakdown of `Good' (coloured sectors) and `Likely' cases (grey sectors) under each subtype.
More than half of the sample are Type Ia SNe (Ia, Ia-91T, and Ia-02cx in the figure).
The second largest group of events are superluminous SNe (SLSNe), which are primarily hydrogen-deficient (SLSN-I) instead of hydrogen-rich (SLSN-II).
The third largest group are `normal' CC SNe that are not classified as superluminous (II, IIP, IIn, Ibn, Ic, and Ic-BL in the figure), among which the fraction of interacting CC SNe (IIn, Ibn) is significantly higher than other comparison samples (Section \ref{sec:samplestat}).
There are also a few SNe with only broad types (`SN' or `SN I').
\label{fig:piechart}}
\end{figure}

\renewcommand{\arraystretch}{1.375}
\begin{table*}
\caption{Number and Fraction by Supernova Type}
\label{tab:typefrac}
\begin{tabular}{@{\extracolsep{6pt}}lccccccc@{}}
\hline
Type    & \multicolumn{2}{c}{Hostless (Primary)} & \multicolumn{2}{c}{Hostless (Extended)} & ZTF BTS                  & TNS                      & LOSS                        \\
          \cline{2-3}                              \cline{4-5}                               \cline{6-6}                \cline{7-7}                \cline{8-8}                 
        & Number & Fraction                      & Number & Fraction                       & Fraction                 & Fraction                 & Fraction                    \\
\hline
Ia      & $76$   & ${60.3}_{-4.8}^{+4.6}\%$      & $100$  & ${62.1}_{-4.2}^{+4.0}\%$       & ${73.7}_{-1.3}^{+1.3}\%$ & ${67.1}_{-0.4}^{+0.4}\%$ & ${79.2}_{-5.5}^{+4.2}\%$    \\
Ibc     & $3$    & ${2.4}_{-1.3}^{+2.3}\%$       & $5$    & ${3.1}_{-1.3}^{+2.0}\%$        & ${6.4}_{-0.7}^{+0.8}\%$  & ${7.3}_{-0.2}^{+0.2}\%$  & ${4.1}_{-1.3}^{+1.6}\%$     \\
II      & $12$   & ${9.5}_{-2.6}^{+3.4}\%$       & $13$   & ${8.1}_{-2.2}^{+2.7}\%$        & ${18.4}_{-1.1}^{+1.2}\%$ & ${24.4}_{-0.4}^{+0.4}\%$ & ${16.6}_{-3.9}^{+5.0}\%$    \\
Ibc+II  & $15$   & ${11.9}_{-2.9}^{+3.6}\%$      & $18$   & ${11.2}_{-2.5}^{+3.0}\%$       & ${24.7}_{-1.3}^{+1.3}\%$ & ${31.8}_{-0.4}^{+0.4}\%$ & ${20.7}_{-4.1}^{+5.2}\%$    \\
Ibn+IIn & $9$    & ${7.1}_{-2.3}^{+3.1}\%$       & $10$   & ${6.2}_{-1.9}^{+2.5}\%$        & ${3.4}_{-0.5}^{+0.6}\%$  & ${3.7}_{-0.2}^{+0.2}\%$  & ${4.8}_{-2.2}^{+2.3}\%$     \\
SLSN    & $33$   & ${26.2}_{-4.1}^{+4.5}\%$      & $39$   & ${24.2}_{-3.5}^{+3.8}\%$       & ${1.6}_{-0.4}^{+0.5}\%$  & ${1.1}_{-0.1}^{+0.1}\%$  & --                          \\
\hline
\end{tabular}
\end{table*}

We show examples of selected hostless SNe in Figure \ref{fig:fieldexamples}. These SNe do not have visible hosts at the sensitivity of the LS.
Even if there are faint sources in the neighbourhood, their $\dDLR$ parameters are above the $\dDLR$ threshold defined in Appendix \ref{appendix:threshold}, large enough to exclude them as possible hosts.

In Figure \ref{fig:piechart}, we outline the breakdown of SN types in our sample. We use the existing classifications in the TNS as reported by individual survey programs.
We also group SNe by their major types and derive their fractions in our sample (Table \ref{tab:typefrac}).
Most SNe in the Extended Sample are SNe Ia, where a few are classified as SN Ia subtypes, including five Ia-91T-like and one Ia-02cx-like (or Iax).
The second-largest major type is superluminous supernova (SLSN), including 33 hydrogen-deficient (SLSN-I) and six hydrogen-rich (SLSN-II) ones.
`Normal' CC SNe, i.e., CC SNe that are not classified as superluminous in TNS, form the third largest group in our sample. There are three II, one IIP, nine IIn, three Ic, one Ic-BL, and one Ibn, which we group together under `Ibc+II' in Table \ref{tab:typefrac}.

To highlight the unique demographics of hostless SNe, we also compile the fractions of major SN types in other SN samples, as summarized in Table \ref{tab:typefrac}.
These samples include: 1) the flux-limited ($<18.5$ mag) ZTF BTS sample \citep{Fremling20, Perley20}; 2) the TNS sample, namely TNS transients with SN classification and discovered after 2016 Jan 1; and 3) the Lick Observatory Supernova Search (LOSS) sample \citep{Li11lumfunc}, where the SN type fractions are estimated for a flux-limited SN survey based on the original galaxy-targeted SN sample.
Most SNe in these comparison samples have host galaxies. The ZTF BTS sample has a hostless fraction of $\sim2$ per cent\footnote{Estimated using the ZTF BTS online catalogue (\url{https://sites.astro.caltech.edu/ztf/bts/bts.php}, accessed on Oct. 15, 2021), which includes host information.}; the TNS sample, per our selection criteria, also has a hostless fraction of $\sim 2$ per cent (161 hostless SNe out of 8137 SNe).

We find several major differences between our hostless SN sample and the comparison samples.
First, SLSNe are particularly abundant among hostless SNe. Nearly a quarter of all SNe in our sample are SLSNe, in striking contrast to the low fraction of $\sim1.6$ per cent in ZTF BTS and $\sim1.1$ per cent in TNS.
There are also nearly twice as many SLSNe as non-SLSN, `normal' CC SNe in our hostless SN sample. However, in ZTF BTS and TNS samples, `normal' CC SNe outnumber SLSNe by more than an order of magnitude.
Second, the ratio of hydrogen-deficient SLSN-I to hydrogen-rich SLSN-II is also higher for hostless SNe. The ZTF BTS sample includes 11 SLSN-I and eight SLSN-II. The TNS sample contains 95 SLSN-I and 40 SLSN-II.
Our sample, however, contains 33 SLSN-I and six SLSN-II ($84.6\pm5.7$ per cent\footnote{{Uncertainties are $68$ per cent binomial proportion confidence intervals unless otherwise noted.}} SLSN-I in all SLSNe, compared to $57.9\pm11.3$ per cent for ZTF BTS and $70.3\pm3.9$ per cent in the TNS sample).
Third, in `normal' CC SNe, there is a significantly higher fraction of interacting subtypes (Ibn and IIn) in our sample compared to their non-interacting counterparts.
We detect $10$ interacting SNe (IIn and Ibn) out of $18$ non-SLSN, `normal' CC SNe (`Ibc+II' in Table \ref{tab:typefrac}), a fraction of $55.6\pm11.6$ per cent interacting SNe in `normal' CC SNe, higher than the fractions of $13.6\pm2.0$ per cent in the ZTF BTS and $11.6\pm0.5$ per cent in the TNS sample.

The high fraction of SLSNe in hostless SNe and hostless CC SNe is largely attributable to a joint selection effect.
We select detectable SNe with undetectable host galaxies, which by nature favours cases with greater contrasts of SN to host luminosities.
SLSNe span a peak absolute magnitude range from $-20$ to $-22$ \citep{GalYam19}, comparable to $L_{\star}$ galaxies and can easily outshine fainter host galaxies.
While the hosts of SLSN-II span a wide range of properties, SLSN-I clearly prefers low-mass, star-forming, and metal-poor galaxies (e.g., \citealt{Angus16, Perley16, Schulze18}), potentially contributing more cases of extreme SN-to-host luminosity contrasts.
The luminosity-driven joint selection effect here could be a reason for the higher fraction of SLSNe, particularly SLSN-I, among hostless SNe.

Interacting SNe represent another phenomenological subgroup of SNe, where the shock interaction with pre-existing circumstellar material leads to narrow emission lines and may boost the luminosity (see the review of \citealt{Smith17}).
SNe IIn show a wide range of absolute magnitudes, and they may prefer later-type, fainter galaxies (e.g., \citealt{Li11lumfunc}; but see also \citealt{Schulze21}). At a higher contrast of SN to host luminosity, some interacting SNe may also appear to be hostless.
In fact, the mean redshift of interacting SNe in our sample  ($0.126\pm0.022$) is higher than their non-interacting siblings ($0.053\pm0.008$), i.e., CC SNe that are neither superluminous nor interacting.
Therefore, the observed higher fraction of interacting SNe could also be attributed to a luminosity-driven joint selection effect.

More generally, for any combination of SN and host luminosities ($L_\mathrm{SN}$, $L_\mathrm{host}$), based on the sensitivity limits of the transient and galaxy surveys, we can derive the maximal luminosity distances ($D_\mathrm{L,SN}$, $D_\mathrm{L,host}$) at which the SN or host becomes undetectable.
The SN will appear hostless if $D_\mathrm{L,SN}>D_\mathrm{L,host}$ \textit{and} if the SN lies at a luminosity distance in-between $D_\mathrm{L,SN}$ and $D_\mathrm{L,host}$. 
SLSNe have greater $D_\mathrm{L,SN}$ on average than other transients due to their higher luminosities, leading to potentially more cases with $D_\mathrm{L,SN}>D_\mathrm{L,host}$.
In addition, the hosts of SLSN-I are usually dwarf galaxies with smaller $D_\mathrm{L,host}$ on average, further expanding the cosmic volume in which the SN may appear hostless.
SNe of lower luminosities (e.g., II-P) have smaller $D_\mathrm{L,SN}$, limiting the potential number of hostless cases. To appear hostless within this distance limit, their hosts must be even fainter in luminosity to satisfy $D_\mathrm{L,host} < D_\mathrm{L,SN}$.
Such faint hosts may not contribute a large number of SNe in a cosmic volume, further limiting the number of hostless cases for these SN types.

Overall, the demographic pattern of our hostless SNe could be attributed to an over-representation of some subtypes (particularly SLSN-I) and a possible deficiency of other low-luminosity subtypes.
However, the discussion here assumes a flux-limited parent SN sample and galaxy catalogue.
In reality, the parent SN sample has contributions from multiple transient surveys with different sensitivity limits; there are also other subtle selection effects in SN detection and classification. These factors may affect the demographic pattern of hostless SNe here.

Finally, the existence of hostless CC SNe disfavours the scenario of hypervelocity stars (HVS).
Typical velocity of Galactic HVS is $1000$ $\text{km}\,\text{s}^{-1}$ \citep{Brown15_araa} or about $1$ $\text{kpc}\,\text{Myr}^{-1}$, while the lifetime of CC SN progenitor is below $50$ Myr for single stars, and no more than $200$ Myr for binary systems \citep{Zapartas17}.
Their progenitors, if dynamically ejected from nearby galaxies soon after formation, travel less than $200$ kpc before the explosion.
Hosts at such distances can be identified during our selection and inspection process, as we find possible unreported hosts at such distances.

\subsection{Constraints on host properties}
\label{sec:limits}

Assuming that the host galaxies of these SNe, if the hosts do exist, are located near the SN sky coordinates and merely lie below the local sensitivity limits of the galaxy survey, we can constrain some properties of the host from the archival multi-band images and stellar population synthesis models.

\subsubsection{Host flux limits}

To begin, we estimate the sensitivity limits of the archival survey images at the SN positions. We use the same images that we earlier inspected to confirm the hostless SNe.
In other words, we estimate the sensitivity limits using the LS images, except when they are contaminated by SN light or do not cover the SN position. For the latter cases, we estimate the sensitivity limits using PS1 images.

The LS data release contains sensitivity maps for point and extended sources, where a circular exponential profile with a $0.45$ arcsec effective radius is used to represent typical extended sources near the detection limits.
For hostless SNe identified with the LS catalogue and images, we employ the corresponding extended source sensitivity map (`galdepth') to estimate the local 5$\sigma$ detection limits.

The PS1 images and catalogue do not have local estimates of sensitivity. There are survey-wide estimates for point source limiting magnitudes, but the extended source limits are likely brighter, and field-to-field variation is non-negligible.
Therefore, we use simple aperture photometry to estimate the sensitivity limits of the PS1 images; we calculate the error of photometric flux inside a $5$ arcsec diameter aperture at the SN position using the variance map and then derive the corresponding local sensitivity limits in magnitudes.
Notably, most of the 56 SNe with $3\sigma$ flux excesses in pre-explosion images, as discussed in Section \ref{sec:samplesummary}, are recovered in PS1 images, primarily in the $i$ and $r$ bands (22 and 13 cases, respectively).

Aperture-based estimates of the PS1 sensitivity limits may deviate from the actual limit of the PS1 photometry pipeline.
PS1 detects faint sources using the signal-to-noise arrays of PSF-convolved images, which is a matched filter technique and favours point or compact sources.
Fixed-diameter aperture photometry, on the other hand, does not assume source morphology and is only sensitive to the integrated aperture flux.
For point or compact sources, our aperture-based limiting magnitudes are more conservative (i.e., more shallow) than the pipeline limiting magnitudes of the PS1 catalogue.
However, for diffuse extended sources that do not exactly match the PSF kernel, our aperture-based limiting magnitudes are likely close to, or even deeper than the pipeline detection limits.

\begin{figure}
\centering
\includegraphics[width=\linewidth]{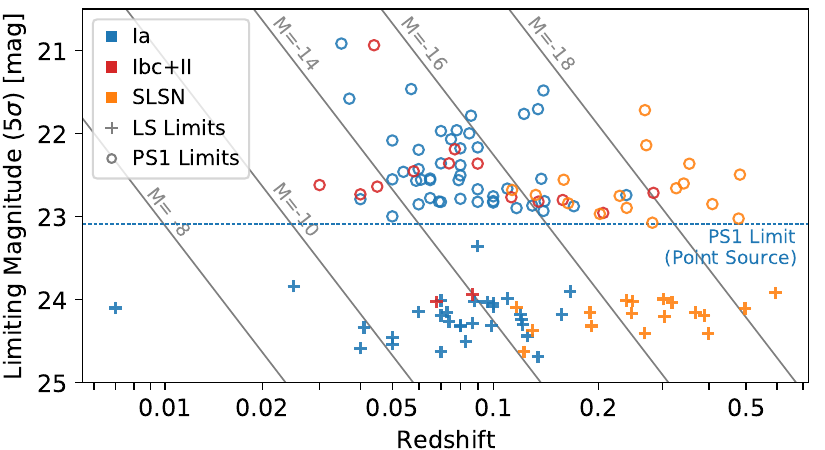}
\caption{
Redshift versus extinction-corrected limiting magnitude ($5\sigma$, $g$-band) for undetected host galaxies of SNe in our Extended Sample.
Limiting magnitudes derived using PS1 and LS images are plotted with different point types, with colours indicating the different SN classes.
We use $5$ arcsec diameter apertures to estimate the limiting magnitude of the PS1 images; the blue horizontal line indicates the $g$-band limit from \citet{Chambers16} corrected for the median value of Galactic foreground extinction at the SN sky coordinates. 
Limiting magnitudes of the LS images are derived using the sensitivity maps for extended sources (`galdepth').
Slanted lines indicate the implied pseudo-absolute magnitude limits, i.e., the absolute magnitudes \textit{without} k-correction.
LS images, on average, provide deeper sensitivity limits than PS1 images.
\label{fig:appmaglimits}}
\end{figure}

We correct the limits for Galactic foreground extinction using the dust map of \citet{Schlegel98} and \citet{Schlafly11}. Extinction coefficients for the LS filters are provided in the DR9 online documentation\footnote{\url{https://www.legacysurvey.org/dr9/catalogs/}}, while coefficients for the PS1 filters are based on \citet{Tonry12}.
The extinction-corrected, $g$-band $5\sigma$ flux limits at the SN sky coordinates versus SN redshifts are shown in Figure \ref{fig:appmaglimits}.
The slanted lines indicate the implied limits of host pseudo absolute magnitudes, i.e., estimates of absolute magnitudes \textit{without} k-correction.

The local sensitivity limits based on the LS images are, on average, deeper than those based on the PS1 images for SNe in our sample. Besides the higher sensitivity of the LS images, another reason is that sensitivity limits for SNe at lower Galactic latitudes, where foreground extinction is higher, are only available from the PS1 images.
The aperture-based limiting magnitudes of the PS1 images are brighter by about half a magnitude than the survey-wide point source limits. Most of our SNe are around $z \sim 0.1$, with SLSNe dominating at high redshift.
Because SN redshifts and types are similar across the LS and PS1-derived limits, we combine these two surveys in the following analyses and do not specify the image sources of the limiting magnitudes.

\subsubsection{Host absolute magnitude limits} \label{sec:absmaglimits}

We derive the implied luminosity limits of possible hosts based on the per-band flux upper limits at the SN positions.
Generally, converting the observed flux to intrinsic luminosity or absolute magnitude requires not just a luminosity distance and dust extinction correction, but also a k-correction (e.g., \citealt{Hogg02}).
K-correction derives the restframe colours and absolute magnitudes using the observed fluxes in redshifted bands and an inferred or assumed intrinsic spectral energy distribution (SED).
However, we have only the flux limits of the hosts, and we do not know their intrinsic SEDs.
Therefore, instead of using the conventional k-correction technique, we \textit{assume} two possible host SEDs and then derive the luminosity limits in the rest-frame $g$-band using the estimated flux limits in multiple bands for each SED.

We choose two simple stellar population (SSP) models for the galaxy SEDs:  `blue' with a stellar population age of $t_{\text{SSP}}=10\,\text{Myr}$, and `red' with $t_{\text{SSP}}=10\,\text{Gyr}$.
These two SEDs are constructed using \textsc{FSPS} \citep{Conroy09, Conroy10} with MIST isochrones \citep{Choi16} and the MILES spectral library \citep{FalconBarroso11}, assuming a \citet{Kroupa01} initial mass function and solar metallicity. Nebular emission, dust attenuation, and thermal re-emission are ignored.
The blue and red SEDs, therefore, represent two extreme cases of spectral shapes in optical wavelengths.
For each SED, we calculate the implied flux in the LS and PS1 filters as a function of redshift.

For each set of observed flux limits (either in the LS or PS1 filters), we scale up the rest-frame $g$-band luminosity of the SED until the implied flux in \textit{any} observed band reaches the extinction-corrected $5\sigma$ upper limit of the real images, i.e., constrained by the sensitivity limits in this band.
The scaled rest-frame $g$-band luminosity is then considered the luminosity upper limit.
We perform the estimate separately using the blue and red SEDs; in other words, each host has two rest-frame $g$-band luminosity upper limits, one assumes the blue SED, and the other assumes the red SED.
The luminosity limits assuming the blue SED are usually constrained by the sensitivity limits of the observed $g$-band (about $90$ per cent cases), while luminosity limits assuming the red SED are usually constrained by the observed $r$-band ($68$ per cent cases) or $z$-band ($28$ per cent cases).

\begin{figure}
\centering
\includegraphics[width=\linewidth]{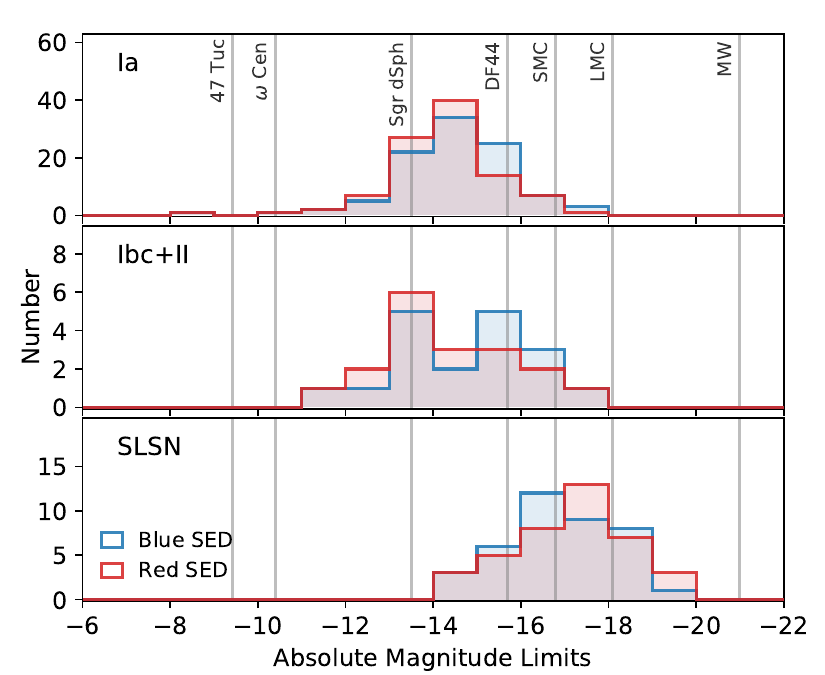}
\caption{
Limiting absolute magnitude distributions in the rest-frame $g$-band for the undetected host galaxies.
We estimate these limits from the $5\sigma$ flux upper limits at the SN coordinates, with Galactic extinction corrections.
Assuming either a `blue' or `red' host spectral energy distribution, the limiting absolute magnitudes are derived based on the local sensitivity limits at SN positions (Section \ref{sec:absmaglimits}).
For comparison, we also plot the absolute magnitudes of example low-luminosity or low-surface brightness objects (gray vertical lines);
their absolute magnitudes are measured in the $V$-band, between the effective wavelengths of the $g$- and $r$-bands of the LS and PS1 filter sets.
DF44 is the exception; its absolute magnitude is reported in the \textit{F606W} band, close to the $r$-band.
\label{fig:absmaglimits}}
\end{figure}

Figure \ref{fig:absmaglimits} shows the derived rest-frame $g$-band $5\sigma$ absolute magnitude limits for the possible host galaxies. 
We compare the blue and red SED absolute magnitude limits for three major SN types: Ia, `normal' CC SN (Ib, Ic, and II; `Ibc+II'), and SLSN.
We also plot the absolute magnitudes of some low-luminosity or low-surface brightness objects as a reference\footnote{We use the \citet{McConnachie12} catalogue for the Milky Way, Magellanic Clouds, and Sagittarius Dwarf Spheroidal. The value for Dragonfly 44 is from \citet{vanDokkum15}. The globular clusters are based on \citet{Harris96} values.}.
Interestingly, the host luminosity limits are comparable to local low-luminosity or low-surface brightness systems; fainter ones are comparable to dwarf spheroidal galaxies (dSph) and even globular clusters.
Assuming the red SED yields fainter limits in the $g$-band, except for SLSNe, whose higher redshifts make the blue SED a more conservative constraint.
Their higher redshifts also lead to brighter luminosity limits for their hosts, although these limits are still comparable to local dwarf galaxies like the Magellanic Clouds.

Using the more conservative (brighter) estimate from either the red or blue SED, we identify at least seven SNe with host absolute magnitudes fainter than $M_g=-12$ (about $1$ per cent the luminosity of the Small Magellanic Cloud), where the faintest reaches $M_g\simeq-8$, close to the luminosity of globular clusters \citep{Harris96} or commonly-defined ultra-faint dwarfs \citep{Simon19}.
Notably, the faintest luminosity limits of ground-based wide-area sky surveys are comparable to \emph{HST}'s host luminosity limits for intracluster SNe \citep{Graham15}, 
demonstrating the effectiveness of those wide-area  surveys in detecting faint hosts or in placing rigorous limits on host luminosities, at least for low-redshift SNe.

\subsubsection{Host stellar mass and star formation rate limits}

The rest-frame $g$-band luminosity limits above are obtained from the $5\sigma$ flux limits in the observed LS or PS1 bands.
Such luminosity limits only rely on the sensitivity limits in a single band and do not directly reflect the physical properties of the possible host galaxies.
To constrain the host properties using \textit{multi-band} data, we perform forced aperture photometry at each SN position and then fit stellar population synthesis (SPS) models to constrain host stellar mass and star formation rate (SFR). Most undetected hosts are consistent with zero aperture flux at the SN position.

We measure the integrated flux in a $5$ arcsec diameter aperture at the sky coordinate of each SN. For the hostless SNe confirmed with LS images, we measure the flux in the LS \textit{grz}-band images; for hostless SNe confirmed with PS1 images, we measure the flux in the PS1 \textit{grizy} images.
When estimating the PS1 sensitivity limits, we assume that the images are properly calibrated and background-subtracted. However, any non-zero local background level may contribute significantly to the measured aperture flux, especially at lower flux levels.
Therefore, we measure the local background using a concentric annular aperture with an inner diameter of $7.5$ arcsec and an outer diameter of $9$ arcsec. The measured background is then scaled by the ratio of aperture areas and subtracted from the inner aperture. Due to this background subtraction process, the uncertainty of the inner aperture flux is higher.
The measured fluxes are also corrected for Galactic foreground extinction using the same procedure as in Section \ref{sec:absmaglimits}.

At a redshift of $0.05$, a diameter of $5$ arcsec corresponds to $4.89$ kpc. At a redshift of $0.1$, $5$ arcsec corresponds to $9.22$ kpc. Both sizes are larger than typical faint dwarf galaxies.
Therefore, our fixed aperture size would measure a dwarf's integrated flux, and local background subtraction is less likely to underestimate the measured flux.

\begin{figure}
\centering
\includegraphics[width=\linewidth]{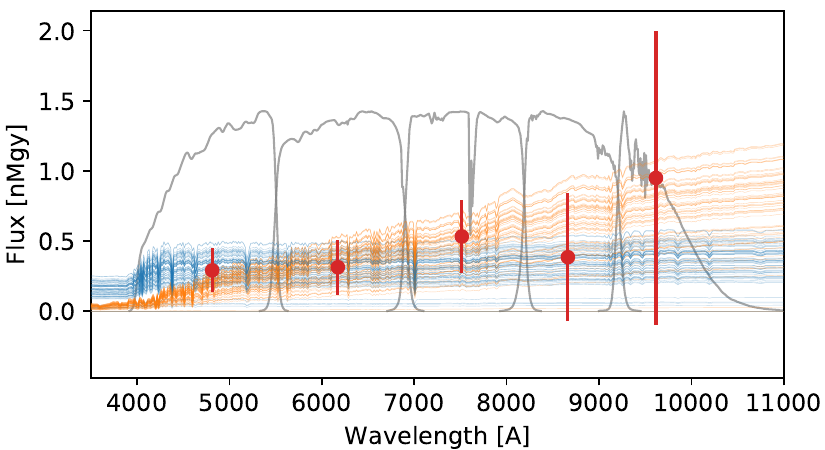}
\caption{
Constraining host stellar mass and star formation rate (SFR) from forced aperture photometry. We use SN 2016iet as an example here.
We first determine extinction-corrected, background-subtracted fluxes within a $5$ arcsec aperture (red points) in either the PS1 or LS filters (gray curves, PS1 \textit{grizy} in this case). Most undetected hosts are consistent with zero aperture flux at the SN position.
We fit the forced photometry fluxes with two stellar population synthesis (SPS) models: one with a constant SFR (blue spectra) and the other with a single, $10$ Gyr stellar population (orange spectra).
Each SPS model only has one free parameter, the total formed mass, for which we use a log-uniform prior in the Monte-Carlo sampling.
We use the single-sided $95$ per cent confidence interval of the posterior to derive the upper limits of host stellar mass and SFR.
Spectra in this figure are randomly drawn from the posterior.
\label{fig:hostsedfit}}
\end{figure}

We use the SPS code \textsc{prospector} \citep{Johnson21} and \textsc{fsps} \citep{Conroy09} to carry out the analysis. Given the low signal-to-noise ratio and limited wavelength coverage, we choose a simple model with constant SFR to represent a star-forming galaxy and a simple stellar population with an age of $t_{\text{SSP}}=10$ Gyr to mimic a quiescent galaxy.
Rather than reusing the blue SED in Section \ref{sec:absmaglimits} to represent star-forming galaxies, we choose a continuous star-forming model here to constrain the current SFR of the hosts.
For either `galaxy,' the only free parameter is the formed stellar mass, which has a log-uniform prior function.
The formed stellar mass here is the direct integration of the star formation history over time, without correcting for the decreasing of remaining stellar mass due to stellar deaths.
Similarly, we do not consider nebular emission, dust attenuation, or thermal re-emission.
We run Markov Chain Monte Carlo (MCMC) sampling for $8192$ steps, removing the first $20$ per cent burn-in phase and using the remaining $80$ per cent as the posterior.
Figure \ref{fig:hostsedfit} shows an example of fitting the aperture fluxes of one SN with the two SPS models.

We then convert the total formed stellar mass from the posterior sample to the present-day stellar mass, where the mass fraction of survival stars and remnants is calculated using \textsc{fsps}.
For the star-forming model, we also divide the formed mass by the age of the universe at the supernova redshift to estimate the current SFR limits.
Assuming a constant SFR, we may overestimate the present-day stellar mass while underestimating the ongoing SFR, \textit{if} these hosts feature a rising star formation history.
Nevertheless, the mass and SFR estimates are only approximations of the diverse and complex star formation histories of galaxies, especially dwarf galaxies.

\begin{figure}
\centering
\includegraphics[width=\linewidth]{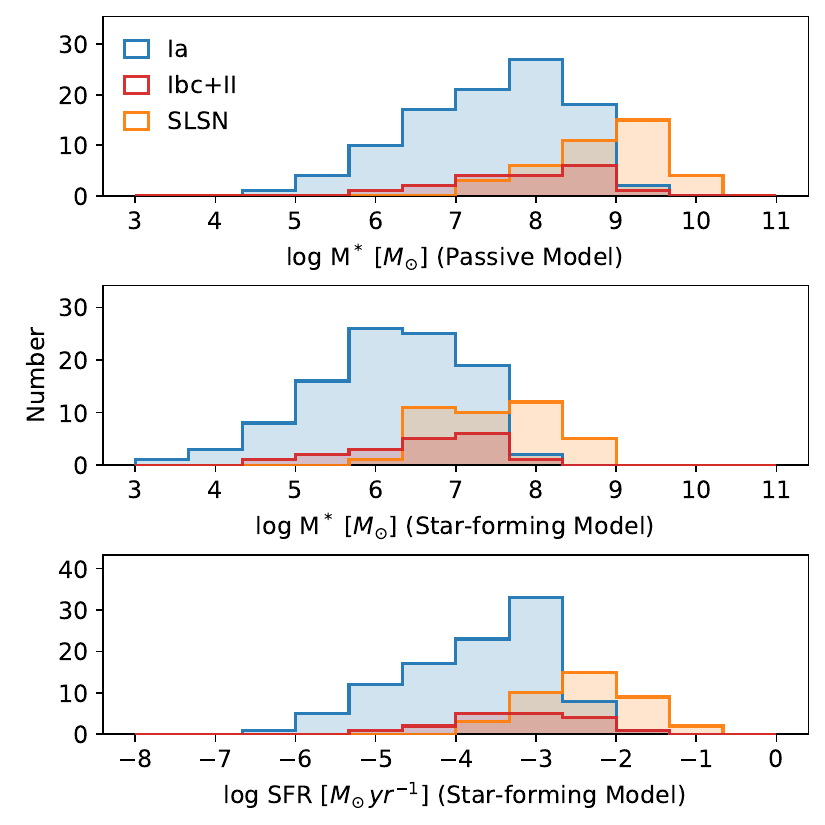}
\caption{
Distributions of upper limits (95th percentile one-sided confidence intervals) on stellar masses ($M^*$) and star formation rates (SFR) from forced aperture photometry.
We estimate the present-day stellar mass from both a quiescent model with a single burst of star formation and a star-forming model with a constant SFR (upper two panels). The star-forming model also places constraints on the current SFR (bottom panel).
Compared to the quiescent model, the star-forming model leads to lower mass upper limits due to the lower mass-to-light ratio of younger stellar populations.
Overall, the stellar mass and SFR limits of SN Ia hosts are comparable to the hosts of SN Ib, Ic, and II. 
The mass and SFR limits of SLSN hosts are higher than other SN types due to the higher SLSN redshifts.
\label{fig:galproplimits}}
\end{figure}

Figure \ref{fig:galproplimits} shows the distribution of stellar mass ($M^*$) and SFR limits estimated from the forced aperture photometry and SED modelling.
To represent the $M^*$ and SFR upper limits, we show the single-sided $95$ per cent confidence interval of the posterior, as higher percentages may suffer from uncertainties due to the limited Monte Carlo sample size.
Compared to the quiescent model, the star-forming model features lower mass-to-light ratios and hence lower $M^*$ upper limits by about $1$ dex.
Assuming quiescent hosts, there are 13 hosts with $M^*<10^6$ $\mathrm{M}_\odot$; assuming star-forming hosts, there are 11 hosts with $M^*<10^5$ $\mathrm{M}_\odot$.
Also, the Ibc+II hosts (CC SNe excluding SLSNe) have higher $M^*$ and SFR upper limits compared with those of SN Ia ($p$-value of $0.02$ from a Student's t-test), although the difference could be driven by the different redshift distributions of subtypes.

Hostless SNe are promising tracers of faint dwarf galaxies \citep{Conroy15, Sedgwick19}; the stellar masses of our unseen host galaxies are generally at the low-mass end of the galaxy stellar mass function.
Our work also suggests that hostless SNe might be used to constrain the integrated $M^*$ and SFR beyond the sensitivity limits of wide-area galaxy surveys.
Improving our analysis requires a large, homogeneous, and unbiased SN sample, which may become available in the near future, particularly with Rubin Observatory's Legacy Survey of Space and Time \citet{Ivezic19}.
Careful calibration of biasing factors, such as the dependence of SN rates on galaxy properties
(e.g., \citealt{Li11rates}; \citealt{Graur17}), should be considered.

\subsection{Testing the association with intracluster starlight}
\label{sec:cluster}

\begin{figure}
\centering
\includegraphics[width=\linewidth]{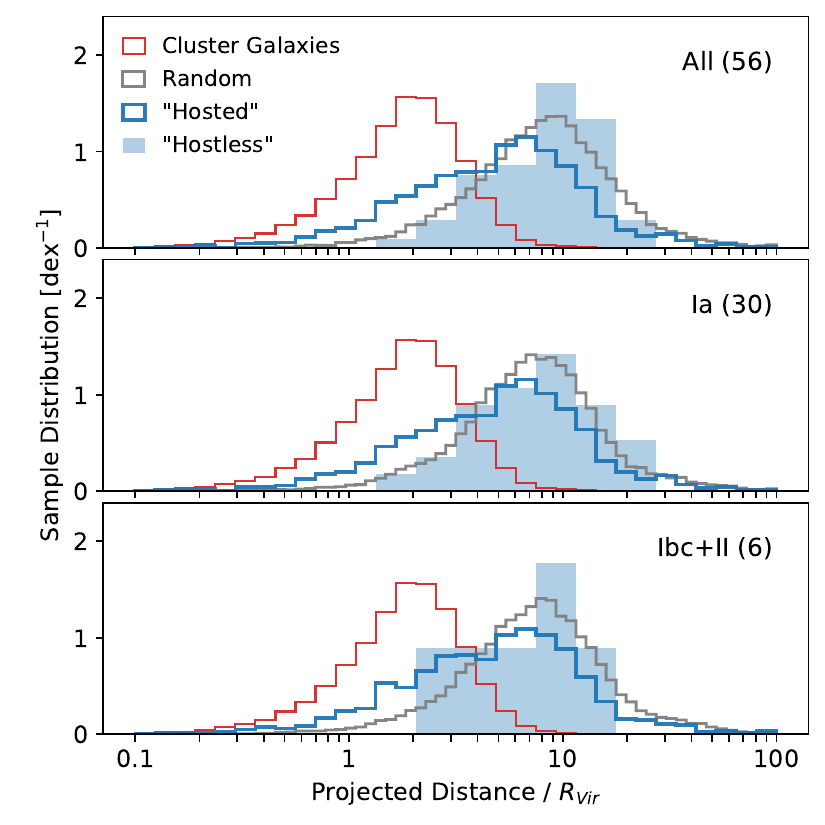}
\caption{
Normalized projected distances from the nearest galaxy group or cluster for hostless SNe (blue shaded), group or cluster member galaxies (red), `hosted' SNe (blue), and mock `field' SNe with randomized sky coordinates (grey).
Besides the full sample (`All'; including SLSNe; top panel), we examine the SNe Ia and `normal' CC SNe (`Ibc+II'; excluding SLSNe) subsamples (bottom two panels).
Hostless SNe are statistically indistinguishable (at greater than $95$ per cent confidence) from the mock SNe in all three panels, suggesting that they originate mostly from field galaxies.
`Hosted' SNe shows stronger clustering than hostless SNe, with more overlap with group or cluster member galaxies.
\label{fig:clusterdistance2}}
\end{figure}

A substantial fraction of the stellar light in galaxy clusters is associated with the diffuse intracluster light (ICL). Depending on the cluster mass and criteria, the fraction varies from a few per cent to about half (Section \ref{sec:intro}).
Hostless SNe associated with ICL are discovered in targeted searches (e.g., \citealt{GalYam03, McGee10, Sand11, Graham15}). Therefore, some hostless SNe in our sample could also be associated with the ICL, a scenario to be tested using their distribution with respect to galaxy clusters.

SNe may also explode in faint galaxies beyond our survey sensitivity limits. Given the correlation of galaxy properties with the local galaxy density (e.g., \citealt{Dressler80}), the clustering (or not) of hostless SNe could be used to constrain the nature of their host galaxies, if they exist.

Therefore, we examine the relationship of hostless SNe to cluster environment and to the more rarefied field.
We find the nearest galaxy group or cluster to each hostless SN and compare the projected distance distribution to that of cluster galaxies, a `field' sample of randomly distributed mock SNe, and a `hosted' SN sample.

\subsubsection{Finding the nearest galaxy clusters}

To identify the nearest galaxy group or cluster to each SN, we choose the SDSS-based galaxy group and cluster catalogue of \citet{Tempel14}.
The catalogue here is constructed with a friends-of-friends algorithm from SDSS spectroscopic galaxies, which effectively covers the redshift interval of our hostless SNe.
The distances to groups or clusters are calculated separately for the angular (transverse) and recessional velocity (radial) directions from each SN, in normalized units.
We normalize the angular distance by the projected virial radius of the system on the sky; we normalize the recessional velocity difference between the SN ($v_\text{SN}$) and system ($v_\text{cl}$) as
\begin{equation}
\Delta v = (v_\text{SN}-v_\text{cl}) / \sqrt{\sigma_v^2 + c^2\delta_{\text{z,SN}}^2},
\end{equation}
where $\Delta v$ is the normalized distance in the velocity space, $\sigma_v$ is the velocity dispersion of the group or cluster, and  $\delta_{\text{z,SN}}$ is the typical SN spectroscopic redshift error.
Because SN spectra are often low-resolution ($R$ of hundreds) with some very broad features (thousands of $\text{km}\,\text{s}^{-1}$), the SN redshift errors are usually higher than the group-cluster velocity dispersion and thus non-negligible.
Our assumption of $\delta_\text{z,SN}=0.008$ is based on the representative error of template-matching algorithms using SN-only templates \citep{Blondin07}.
We assume non-relativistic recessional velocity since relativistic and non-relativistic recessional velocities differ by only a few per cent in the redshift range of the \citet{Tempel14} catalogue ($z<0.2$).
We then identify the group or cluster with the minimal quadratic sum of normalized distances in the angular and redshift directions to each SN. For robustness, groups or clusters with less than three members are excluded.

\subsubsection{Reference samples}

We compare our hostless SNe to 1) a sample of `hosted' SNe, namely SNe with reported hosts and those are excluded during our sample selection; 2) a mock SN sample, which follows the redshift distribution of hostless SNe, but with coordinates randomized over the sky coverage of the \citet{Tempel14} catalogue; 3) group or cluster member galaxies in the \citet{Tempel14} catalogue.

The `hosted' SN sample is constructed from TNS transients that have SN classifications, lie within the sky coverage of the LS or PS1, and are not in our hostless sample. Because the hosted redshifts are a mix of host galaxy and SN redshifts, we continue to employ the SN redshift error term for distance normalization.
Some `hosted' SNe with host galaxy-based redshifts may thus have lower $\Delta v$.

The mock SN sample represents the scenario where SNe are randomly distributed over the sky, without a statistical association with groups and clusters.
We generate the mock SNe by resampling the redshifts of hostless SNe with replacement and then assigning them random sky coordinates with uniform density on the celestial sphere, over the coverage of the \citet{Tempel14} catalogue.
For a robust comparison, we generate $16384$ mock SNe and assume that their projected distance distribution is exact.

Finally, we create a reference sample of group and cluster member galaxies that represents SNe associated with group or cluster environments.
For consistency and as a test of our method, we use the same procedure to find the nearest group or cluster to each of these member galaxies.
When finding their nearest groups or clusters, we do not include the SN redshift error term for the normalization of $\Delta v$ because cluster member redshifts are precise.
Our method correctly recovered the group or cluster each member galaxy was assigned to in \citet{Tempel14}.

\subsubsection{Comparing hostless SNe to field and intracluster environments}

We compare only the normalized angular distance of hostless SNe with the reference samples, because it should not be affected by the SN redshift errors.
Figure \ref{fig:clusterdistance2} shows the distances to the nearest group or cluster that we identified.
Hostless SNe typically lie many virial radii from the centre, further than the known cluster or group member galaxies.
As expected for a randomized sample, the mock SNe are similarly far from the centres; a Kolmogorov-Smirnov (K-S) test does not distinguish between the mock and hostless SN samples, while distinguishing both from the member sample at $>2\sigma$ confidence.
A K-S test shows that the
hosted SNe are differently distributed than the hostless sample (except for the smaller `Ibc+II' sample), having more overlap with the cluster member galaxies and extending to within a virial radius of the centre.
%

The results here indicate that the hostless SNe in our sample are not generally associated with group or cluster environments and so are unlikely to originate from the diffuse intracluster light. Their host galaxies, if they exist, are most likely faint field galaxies.
The hosted SNe, in contrast, occur not just in the field, but also in galaxies that are close, or even bound to clusters or groups. Therefore, on average, hosted SNe are closer to group or cluster centres than hostless SNe.

Notably, due to the conservative $\dDLR$ threshold we choose (Appendix \ref{appendix:threshold}), we may have undercounted hostless SNe that originate in the ICL where galaxy density is also higher.
We also emphasize that our comparison above cannot exclude a possible contribution from intergalactic stars beyond groups and clusters. 
ICL and `field' intergalactic stars collectively contribute to the Extragalactic Background Light \citep{Cooray12} and may account for about half of stellar light in the universe \citep{Zemcov14}.
The question of whether there are any SNe associated with `field' intergalactic stars remains to be investigated.
Furthermore, the number ratio of cluster to field galaxies in which the hosted SNe lie is not necessarily representative of the number or mass density of galaxies in these two environments; low-mass, star-forming galaxies prefer the field and are expected to produce more SNe per unit stellar mass (e.g., \citealt{Li11, Graur17}).

\subsection{Light curves of hostless Type Ia SNe}
\label{sec:lc}

Type Ia SNe form the single largest class in our sample. Their light curve widths and intrinsic colours correlate with the peak luminosities (e.g., \citealt{Phillips93, Tripp98}), making these SNe standardizable candles at cosmological distances.
Recent cosmological analyses incorporate host properties, primarily the stellar mass (e.g., \citealt{Scolnic18, Smith20des}), as a dependence factor of SN Ia luminosity. Meanwhile, hostless SNe are routinely included in such samples, under the assumption that their hosts are undetected due to their low stellar masses.
It is thus interesting to compare their light curve properties with SNe Ia that occurred in detectable galaxies and to consider the implications in the context of SN-host correlations.

\subsubsection{Light curve data sources}

Most SNe in our sample were discovered by a few transient sky surveys, including ZTF and its predecessor, the Palomar Transient Factory (PTF, iPTF; \citealt{Law09}), the Asteroid Terrestrial-impact Last Alert System (ATLAS; \citealt{Tonry12, Smith20atlas}), \textit{Gaia}, the All-Sky Automated Survey for Supernovae (ASAS-SN), and Pan-STARRS \citep{Tonry12}.
Most of these provide light curve data or portals via which light curves can be constructed on request.
We choose ATLAS and ZTF as our data sources for their adequate sensitivity and pixel scale, easier scripted data access, and more complete light curve information (e.g., quality flags and errors).
%

The ATLAS survey scans the sky northward to $\text{Dec}\sim-50^{\circ}$ in a two-day cadence in two wide-band filters, a `cyan' filter (ATLAS-$c$) and an `orange' filter (ATLAS-$o$), to a single-epoch limiting magnitude of $\sim 19.5$ ($o$-band).
We access ATLAS light curves via their forced photometry service\footnote{\url{https://fallingstar-data.com/forcedphot/}}, using the SN coordinates from the TNS as input.
The light curves are limited to within $120$ days before and $240$ days after the initial discovery date in the TNS. These limits are sufficient to cover the phase range that the \textsc{SALT2} model \citep{Guy07, Guy10} can fit.
We perform forced PSF photometry on difference images, rather than calibrated science exposures, as our targets are transient by nature.
Filter transmission curves for the ATLAS bands are obtained from the Filter Profile Service\footnote{\url{svo2.cab.inta-csic.es/theory/fps/}}.
%

We use publicly-available ZTF images\footnote{\url{https://www.ztf.caltech.edu/ztf-public-releases.html}} to construct light curves.
ZTF surveys the sky northward to $\text{Dec}\sim-30^\circ$ in the $g$, $r$, and $i$ bands, with a single-epoch limiting magnitude of $20.5$ ($r$-band) and a range of cadences depending on the field and science program.
The data release already contains light curves, but they are constructed by matching single-epoch point source catalogues of calibrated science exposures rather than difference images, and are thus not optimal for SNe.
The forced photometry service, despite being available, is not optimized for large volume of queries.
Therefore, we construct light curves by performing forced PSF photometry directly on the difference images. We access postage stamps at the SN sky coordinates from 120 days before to 240 days after the discovery date in the TNS.
The procedure of forced photometry is discussed in Appendix \ref{appendix:psfphotometry}.
Besides the light curve sample of hostless SNe Ia, we also assemble a reference light curve sample of `hosted' SNe Ia with the host galaxy database of \citet{Qin22}. 
To construct it, we choose SNe Ia in the parent TNS sample whose host galaxies are cross-matched with the SDSS Main Galaxy Sample (MGS; \citealt{Strauss02}) spectroscopic targets.
The MGS is a nearly flux-limited galaxy sample ($r\sim17.8$). Therefore, the reference sample, which is a subset of the `hosted' SNe in Section \ref{sec:cluster}, represents SNe Ia with relatively bright host galaxies.
We choose the same light curve data sources (ZTF, ATLAS) as we do for the hostless sample.

\subsubsection{Fitting the light curves}

We choose the \textsc{SALT2} spectrophotometric model \citep{Guy07, Guy10} and \textsc{sncosmo} software package \citep{Barbary16} to fit the light curves.
Two relevant parameters are $x_1$ and $c$. The parameter $x_1$ primarily traces the `stretch' of SN Ia light curves, where a more positive value indicates slowly-evolving (or broader) light curves, and a more negative value indicates fast-evolving (or narrower) light curves. 
This parameter mainly accounts for the classic relationship described by \citet{Phillips93}, where slowly-evolving SNe Ia tend to be more luminous, and fast-evolving SNe Ia tend to be less luminous.
The parameter $c$ is a color parameter that can be influenced by intrinsic properties or due to extinction. A positive value of c indicates a redder SED, while a negative value indicates a bluer SED.
%

We choose the same data quality cut and light curve fitting procedure for the hostless and reference samples.
We select ATLAS epochs with a reduced $\chi^2 < 5$ and without data quality issues (\texttt{err=0}). Epochs with photometric zero-points brighter than $18$ mag are excluded due to possibly unfavourable observing conditions.
We choose ZTF epochs with no quality issues flagged in the mask images. Some reference images in the ZTF data release are contaminated by SN light, and the over-subtraction in difference images leads to a shifted flux baseline in some light curves, especially for SNe observed during 2018 and early 2019.
Therefore, we add a flux baseline parameter (in nanomaggies) to each ZTF band, and they are also free parameters to fit.
When fitting the light curves, we use the dust reddening maps of \citet{Schlegel98} and \citet{Schlafly11} to estimate the Galactic foreground extinction at each SN position.
Despite the absence of detectable host galaxies, we make host galaxy extinction (`\texttt{hostebv},' measured in colour excess) a free parameter for consistent comparison with our reference sample.
Both Galactic foreground and host galaxy dust extinction are based on the \citet{Fitzpatrick99} extinction curve.
The host extinction parameter may absorb the uncertainties in the Galactic extinction and will have degeneracies with $c$. Therefore, our comparison mainly focuses on the more robustly constrained $x_1$.
SN redshifts are fixed to the reported values in the TNS.
The model includes the following parameters: time $t_0$, amplitude $x_0$, $x_1$, $c$, and \texttt{hostebv}, with per-band flux baseline parameters in the ZTF filters.

We use \textsc{emcee} \citep{ForemanMackey13}, a Markov Chain Monte Carlo package, to sample the parameter space. Using twice the number of free parameters (\texttt{N\_dim}) as the chain number, after $200$ burn-in steps, we run $1000$ steps to obtain $2000\times$\texttt{N\_dim} samples.
We use the \textit{max a posteriori} parameter as the best-fitting parameter, and the standard deviation of the marginalized posterior as the uncertainty of each best-fitting parameter.

\subsubsection{Light curve stretch and intrinsic colour}

In both the hostless and the hosted reference samples, we select fitting results with $|x_1|<3$, $|c|<0.3$, uncertainty in $x_1$ $< 0.5$, uncertainty in $c$ $< 0.1$, and uncertainty in $t_0$ $< 2$ days.
Except for the relatively tolerant selection cut on the $c$ uncertainty, these criteria are similar, if not more strict, than those used in previous cosmological analyses (e.g., \citealt{Scolnic18, Smith20des}).
The fitting quality criteria here select $49$ out of $75$ hostless SNe Ia, and $636$ out of $1864$ SNe Ia in the reference sample.

\begin{figure}
\centering
\includegraphics[width=\linewidth]{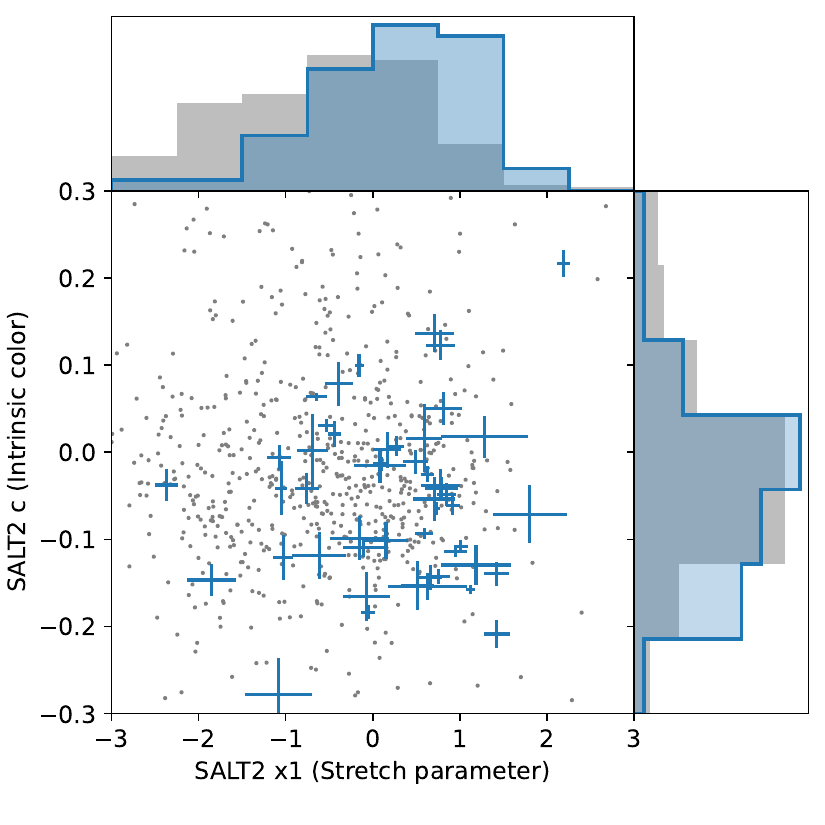}
\caption{Best-fitting \textsc{SALT2} light curve parameters, stretch $x_1$ and intrinsic colour $c$, for hostless SNe Ia (blue), compared with the reference sample of `hosted' SNe Ia (grey) that occurred in galaxies in the flux-limited SDSS Main Galaxy Sample \citep{Strauss02}. The bin widths are set to be three times the median errors in $x_1$ and $c$ (0.249 and 0.086, respectively), across the combined hostless and reference samples.
The hostless sample has a higher fraction of luminous, slowly fading (positive $x_1$) and likely bluer (negative $c$) SNe than the reference sample, a robust conclusion given that we apply the same data sources (ZTF, ATLAS), fitting procedure, and quality criteria for both the hostless and reference samples. The concentration of $x_1$ values above zero for hostless SNe Ia is similar to that observed for star forming hosts in the field in \citet{Larison24} (also see conclusions drawn from Figure \ref{fig:lightcurvegalprop}). 
\label{fig:lightcurveparam}}
\end{figure}

We show the light curve parameters in Figure \ref{fig:lightcurveparam}.
Compared to the hosted reference sample, our hostless SNe lie on the luminous-and-slow (higher $x_1$ value) side of the distribution. Also, their intrinsic colour is likely bluer (lower $c$ value).
The shifts of the mean values are $\Delta\overline{x_1}=0.800\pm0.019$ and $\Delta\overline{c}=-0.0355\pm0.0020$, respectively.
Given that we have selected identical light curve data sources, fitting procedures, and post-fitting quality criteria, it is unlikely that the observed shift is attributable to sample bias.
Given that the intrinsic colour is in degeneracy with \texttt{hostebv} and they both absorb the uncertainties in the Galactic foreground extinction, we only discuss the implications of $x_1$ here.

We measure a positive correlation of redshift with $x_1$ in the reference sample (Pearson's $r=0.130$, $p=0.001$), aligned with the trend that high-stretch SNe are more luminous and thus observable at greater distances.
However, this correlation predicts a mean value of $\overline{x_1}=-0.393\pm0.004$ using the redshifts of hostless SNe, lower than the actual mean value of $\overline{x_1}=0.245\pm0.018$ for the hostless sample.
Therefore, redshift-driven bias may not be the reason for the shift of $x_1$ across the two samples.

\begin{figure}
\centering
\includegraphics[width=\linewidth]{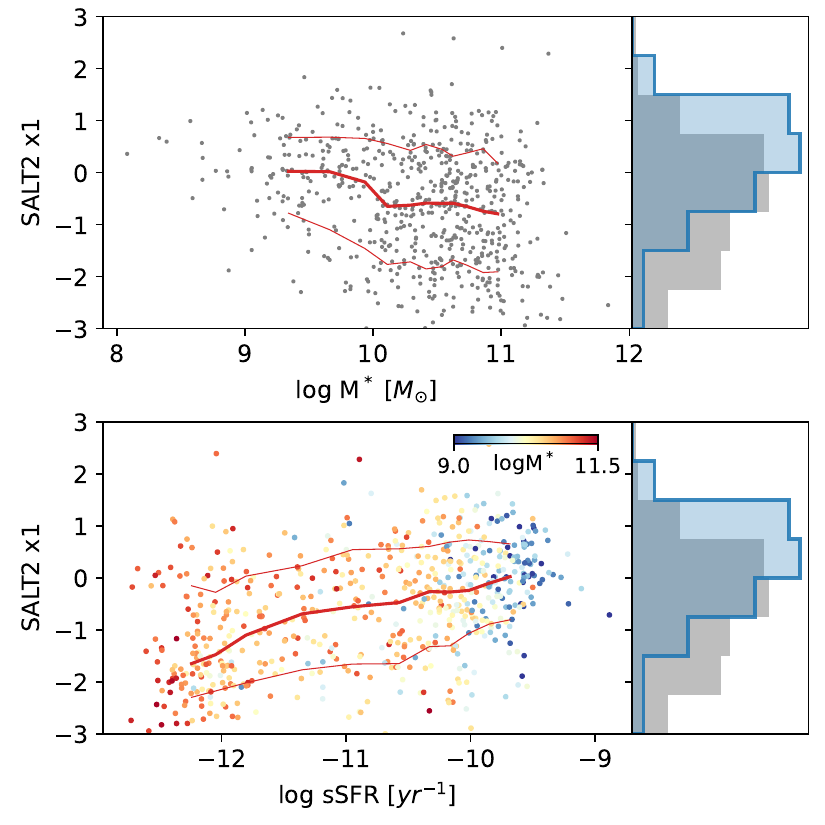}
\caption{
\textsc{SALT2} light curve stretch parameter $x_1$ versus host galaxy stellar mass ($M^*$) and specific star formation rate (sSFR) for the hosted SN Ia sample described in Section \ref{sec:lc}.
Red lines show the median value and the $16$th, $84$th percentiles in a sliding bin of $100$ data points.
The points are coloured in the lower panel by host stellar mass from blue to red, a range corresponding to $9.5<\log M^*<11$.
The righthand panels show the distribution of $x_1$ for this sample (grey histograms)  compared to that of our hostless SN Ia sample (blue histograms).
The hostless SNe Ia are shifted toward the luminous and slowly fading side of the distribution (positive $x_1$).
If the correlations of $x_1$ with $M^*$ and sSFR also hold for the undetected hosts of our hostless sample, then those hosts are likely low-mass and actively star-forming.
\label{fig:lightcurvegalprop}}
\end{figure}

We next consider the shift of the $x_1$ parameter in the context of SN-host correlations.
Multiple SN samples (e.g., \citealt{Sullivan10, Lampeitl10, Childress13, Pan14}) reveal that luminous and slowly-fading SNe Ia (positive $x_1$) prefer low-mass, star-forming (i.e., high specific star formation rate; sSFR) galaxies, while faint and fast-fading SNe Ia (negative $x_1$) are more often seen in high-mass, quiescent galaxies.
We also see such correlations in our reference sample, using the stellar mass and SFR estimates of \citet{Kauffmann03} and \citet{Brinchmann04} (Figure \ref{fig:lightcurvegalprop}).
If such correlations hold for the unseen host galaxies of our hostless SNe Ia sample, then the shift of the $x_1$ parameter indicates that those hosts are likely low-mass, star-forming galaxies.
Compared to the reference sample, the $x_1$ parameter distribution of our hostless SNe Ia sample is shifted to the luminous and slowly fading side, which are preferably hosted by low-mass and high sSFR galaxies.
This indication that they occur in low-mass galaxies is also consistent with our sample selection criteria, which favour luminous transients (e.g., SNe Ia) in low-luminosity hosts.
The current SN Ia rate in a galaxy is connected to its formation history via the delay time distribution (DTD), i.e., the time evolution of the SN Ia rate per unit formed stellar mass after an instantaneous burst of star formation (see the review of \citealt{Maoz12}).
Previous studies revealed that a combination of down-sizing galaxy formation history and a common power-law decay DTD model produces a strong dependence of SN Ia progenitor age distribution on host stellar mass \citep{Childress14}.
Low-mass galaxies, on average, have younger SN Ia progenitors that are contributed by recent star formation ($<1$ Gyr); in contrast, massive galaxies are usually quiescent, and their SN Ia populations are more associated with old stellar populations and longer time delays (several Gyrs).
Combined with the relationship between stretch parameter and progenitor age (e.g., \citealt{Brandt10}), such a progenitor age--host stellar mass relationship naturally predicts the correlation of $x_1$ with host stellar mass and sSFR.

The correlation of $x_1$ with stellar mass and sSFR allows us to infer the properties of the undetected hosts. Given the higher average $x_1$ of hostless SNe Ia compared to `hosted' SNe Ia, it is possible that the undetected hosts of our SNe Ia sample, if they exist, are star-forming dwarf galaxies in which SN Ia progenitors are younger (i.e., with shorter delay times).
Such galaxies could also host CC SNe, whose progenitors also have short delay times ($<200$ Myr; e.g., \citealt{Zapartas17}), observed in our sample.
Alternatively, their unseen, low-luminosity host galaxies could be quiescent galaxies. However, such quiescent dwarf galaxies are usually environmentally-quenched and are thus related to cluster environments, in contrast to our conclusion in Section \ref{sec:cluster}.

\section{Conclusions}
\label{sec:summ}

In this work, we identify a sample of $161$ hostless SNe (including $126$ robust and $35$ likely ones) from public transient records reported in the Transient Name Server since early 2016.
We improve the criteria for host association and choose conservative distance thresholds so that SNe with previously unreported host galaxies in the Pan-STARRS and DESI Legacy Surveys catalogues can be properly excluded.
These SNe are then visually inspected using the images from the two surveys to ensure that there are no visible host galaxies inside or beyond our search radius for catalogue sources.

There are a wide variety of SN types in our hostless SN sample. Both SNe Ia and CC SNe are present, suggesting a population of faint host galaxies that are currently star-forming.
In contrast to the flux-limited SN sample of ZTF BTS \citep{Fremling20, Perley16}, we identify a significantly higher fraction of SLSNe ($24.3_{-2.5}^{+3.8}$ per cent in our hostless sample vs. $1.6_{-0.4}^{+0.5}$ per cent in ZTF BTS), in which hydrogen-deficient SLSN-I outnumber hydrogen-rich SLSN-II (33 SLSN-I vs. 6 SLSN-II, or $84.6\pm5.7$ per cent SLSN-I in all SLSNe, compared to $57.9\pm11.3$ per cent in the ZTF BTS).
The higher fraction of SLSNe than other subtypes in our sample is attributable to the higher intrinsic luminosities of SLSNe and the preference of SLSN-I for low-mass dwarf galaxies.
We also identify a higher fraction of interacting SNe (Type IIn and Ibn) within non-superluminous, `normal' CC SNe ($55.6_{-14.3}^{+13.6}$ per cent interacting SNe in `normal' CC SNe, compared with $13.6\pm2.0$ per cent in the ZTF BTS).
This higher fraction could be a result of their broader luminosity range, where the brightest may outshine their host galaxies.
The prevalence of hostless CC SNe disfavors hypervelocity stars as their progenitors given the velocity of such stars and the delay time of CC SN progenitors.

Using the local sensitivity limits of archival images, we estimate the host luminosity upper limits.
The absolute magnitude limits of the faintest hosts are $M_g\sim-12$, about $1$ per cent the luminosity of the SMC and comparable to dwarf satellite galaxies in the Local Group; the faintest is even close to ultra faint dwarfs or globular clusters.
We also measure the aperture flux at each SN position, with which we constrain the present-day stellar mass and SFR of their hosts using simple stellar population synthesis models: a `quiescent' galaxy with a single burst of star formation at a lookback time of $10$ Gyr and a constant SFR galaxy.
Assuming the quiescent model, there are 13 hosts with $M^*<10^6$ $\mathrm{M}_\odot$; assuming the constant SFR model, there are 11 hosts with $M^*<10^5$ $\mathrm{M}_\odot$.
Our work demonstrates that hostless SNe are effective tracers of stellar mass and star formation beyond the sensitivity limits of wide-area sky surveys. 

To probe the environments of these hostless SNe, we consider their spatial distribution with respect to known galaxy groups and clusters.
We compare the projected separations, normalized by the virial radii, from the nearest group or cluster of our hostless SNe, our hosted SNe, a mock SN sample with randomized locations on the sky, and cluster member galaxies.
Unlike SNe with host galaxies, the projected separation distribution of hostless SNe is indistinguishable from the sample of random mock SNe, suggesting that they
lie in the field instead of higher galaxy density regions.
Therefore, the hostless SNe are unlikely to arise from intracluster starlight; instead, their undetected hosts are more consistent with dwarf galaxies in the field.

We also model the light curves of hostless SNe Ia using archival data from ATLAS and ZTF.
Compared to a similar SN Ia sample with relatively bright host galaxies, we find an excess of luminous and slowly fading SNe Ia in our hostless sample, which cannot be explained by redshift-driven biases.
We also find a possible excess of intrinsically bluer SNe Ia.
If the dependence of light curve parameters on host stellar mass and SFR also holds for the unseen hosts of our hostless SNe Ia, then these results again favour a population of low-luminosity and actively star-forming hosts.

The results on the demographics, cosmic environments, and SN Ia light-curve properties of our hostless SNe favour a population of faint and star-forming host galaxies.

\section*{Acknowledgements}

We are grateful to the anonymous reviewer for their insightful and constructive feedback, which has significantly enhanced the quality of our manuscript.
YQ thank David Sand and Peter Behroozi for constructive comments that helped improving the manuscript. YQ and AIZ acknowledge support from NASA ADAP grant \#80NSSC21K0988. AIZ also thanks the hospitality of the Columbia Astrophysics Laboratory at Columbia University, where some of this work was completed.

The Pan-STARRS1 Surveys (PS1) and the PS1 public science archive have been made possible through contributions by the Institute for Astronomy, the University of Hawaii, the Pan-STARRS Project Office, the Max-Planck Society and its participating institutes, the Max Planck Institute for Astronomy, Heidelberg and the Max Planck Institute for Extraterrestrial Physics, Garching, The Johns Hopkins University, Durham University, the University of Edinburgh, the Queen's University Belfast, the Harvard-Smithsonian Center for Astrophysics, the Las Cumbres Observatory Global Telescope Network Incorporated, the National Central University of Taiwan, the Space Telescope Science Institute, the National Aeronautics and Space Administration under Grant No. NNX08AR22G issued through the Planetary Science Division of the NASA Science Mission Directorate, the National Science Foundation Grant No. AST-1238877, the University of Maryland, Eotvos Lorand University (ELTE), the Los Alamos National Laboratory, and the Gordon and Betty Moore Foundation.

The Legacy Surveys consist of three individual and complementary projects: the Dark Energy Camera Legacy Survey (DECaLS; Proposal ID \#2014B-0404; PIs: David Schlegel and Arjun Dey), the Beijing-Arizona Sky Survey (BASS; NOAO Prop. ID \#2015A-0801; PIs: Zhou Xu and Xiaohui Fan), and the Mayall z-band Legacy Survey (MzLS; Prop. ID \#2016A-0453; PI: Arjun Dey). DECaLS, BASS and MzLS together include data obtained, respectively, at the Blanco telescope, Cerro Tololo Inter-American Observatory, NSF’s NOIRLab; the Bok telescope, Steward Observatory, University of Arizona; and the Mayall telescope, Kitt Peak National Observatory, NOIRLab. The Legacy Surveys project is honoured to be permitted to conduct astronomical research on Iolkam Du’ag (Kitt Peak), a mountain with particular significance to the Tohono O’odham Nation.

NOIRLab is operated by the Association of Universities for Research in Astronomy (AURA) under a cooperative agreement with the National Science Foundation.

This project used data obtained with the Dark Energy Camera (DECam), which was constructed by the Dark Energy Survey (DES) collaboration. Funding for the DES Projects has been provided by the U.S. Department of Energy, the U.S. National Science Foundation, the Ministry of Science and Education of Spain, the Science and Technology Facilities Council of the United Kingdom, the Higher Education Funding Council for England, the National Center for Supercomputing Applications at the University of Illinois at Urbana-Champaign, the Kavli Institute of Cosmological Physics at the University of Chicago, Center for Cosmology and Astro-Particle Physics at the Ohio State University, the Mitchell Institute for Fundamental Physics and Astronomy at Texas A\&M University, Financiadora de Estudos e Projetos, Fundacao Carlos Chagas Filho de Amparo, Financiadora de Estudos e Projetos, Fundacao Carlos Chagas Filho de Amparo a Pesquisa do Estado do Rio de Janeiro, Conselho Nacional de Desenvolvimento Cientifico e Tecnologico and the Ministerio da Ciencia, Tecnologia e Inovacao, the Deutsche Forschungsgemeinschaft and the Collaborating Institutions in the Dark Energy Survey. The Collaborating Institutions are Argonne National Laboratory, the University of California at Santa Cruz, the University of Cambridge, Centro de Investigaciones Energeticas, Medioambientales y Tecnologicas-Madrid, the University of Chicago, University College London, the DES-Brazil Consortium, the University of Edinburgh, the Eidgenossische Technische Hochschule (ETH) Zurich, Fermi National Accelerator Laboratory, the University of Illinois at Urbana-Champaign, the Institut de Ciencies de l’Espai (IEEC/CSIC), the Institut de Fisica d’Altes Energies, Lawrence Berkeley National Laboratory, the Ludwig Maximilians Universitat Munchen and the associated Excellence Cluster Universe, the University of Michigan, NSF’s NOIRLab, the University of Nottingham, the Ohio State University, the University of Pennsylvania, the University of Portsmouth, SLAC National Accelerator Laboratory, Stanford University, the University of Sussex, and Texas A\&M University.

BASS is a key project of the Telescope Access Program (TAP), which has been funded by the National Astronomical Observatories of China, the Chinese Academy of Sciences (the Strategic Priority Research Program “The Emergence of Cosmological Structures” Grant \# XDB09000000), and the Special Fund for Astronomy from the Ministry of Finance. The BASS is also supported by the External Cooperation Program of Chinese Academy of Sciences (Grant \# 114A11KYSB20160057), and Chinese National Natural Science Foundation (Grant \# 11433005).

The Legacy Survey team makes use of data products from the Near-Earth Object Wide-field Infrared Survey Explorer (NEOWISE), which is a project of the Jet Propulsion Laboratory/California Institute of Technology. NEOWISE is funded by the National Aeronautics and Space Administration.

The Legacy Surveys imaging of the DESI footprint is supported by the Director, Office of Science, Office of High Energy Physics of the U.S. Department of Energy under Contract No. DE-AC02-05CH1123, by the National Energy Research Scientific Computing Center, a DOE Office of Science User Facility under the same contract; and by the U.S. National Science Foundation, Division of Astronomical Sciences under Contract No. AST-0950945 to NOAO.

The Photometric Redshifts for the Legacy Surveys (PRLS) catalog used in this paper was produced thanks to funding from the U.S. Department of Energy Office of Science, Office of High Energy Physics via grant DE-SC0007914.

This work has made use of data based on observations obtained with the Samuel Oschin 48-inch Telescope at the Palomar Observatory as part of the Zwicky Transient Facility project. ZTF is supported by the National Science Foundation under Grant No. AST-1440341 and a collaboration including Caltech, IPAC, the Weizmann Institute for Science, the Oskar Klein Center at Stockholm University, the University of Maryland, the University of Washington, Deutsches Elektronen-Synchrotron and Humboldt University, Los Alamos National Laboratories, the TANGO Consortium of Taiwan, the University of Wisconsin at Milwaukee, and Lawrence Berkeley National Laboratories. Operations are conducted by COO, IPAC, and UW. 

This work has made use of data from the Asteroid Terrestrial-impact Last Alert System (ATLAS) project. The Asteroid Terrestrial-impact Last Alert System (ATLAS) project is primarily funded to search for near earth asteroids through NASA grants NN12AR55G, 80NSSC18K0284, and 80NSSC18K1575; byproducts of the NEO search include images and catalogs from the survey area. This work was partially funded by \textit{Kepler}/K2 grant J1944/80NSSC19K0112 and \textit{HST} GO-15889, and STFC grants ST/T000198/1 and ST/S006109/1. The ATLAS science products have been made possible through the contributions of the University of Hawaii Institute for Astronomy, the Queen’s University Belfast, the Space Telescope Science Institute, the South African Astronomical Observatory, and The Millennium Institute of Astrophysics (MAS), Chile.

This material is based upon High Performance Computing (HPC) resources supported by the University of Arizona TRIF, UITS, and Research, Innovation, and Impact (RII) and maintained by the UArizona Research Technologies department.

This research uses services or data provided by the Astro Data Lab at NSF’s NOIRLab. NOIRLab is operated by the Association of Universities for Research in Astronomy (AURA), Inc. under a cooperative agreement with the National Science Foundation.

This research uses the services provided by the Mikulski Archive for Space Telescopes (MAST).

\section*{Data Availability}

The full catalogue of hostless SNe is available as online supplementary material.
This research is based on the publicly-available images and source catalogues of the DESI Legacy Surveys and Pan-STARRS, accessed via Legacy Surveys Sky Viewer (\url{https://www.legacysurvey.org/viewer}), NOIRLab Datalab (\url{https://datalab.noirlab.edu/}), and MAST (\url{https://archive.stsci.edu/}).
The derived data generated in this research will be shared on reasonable request to the corresponding author.



\bibliographystyle{mnras}
\bibliography{main} 




\appendix

\section{Adaptive threshold of $d_{DLR}$}
\label{appendix:threshold}

\begin{figure}
\centering
\includegraphics[width=\linewidth]{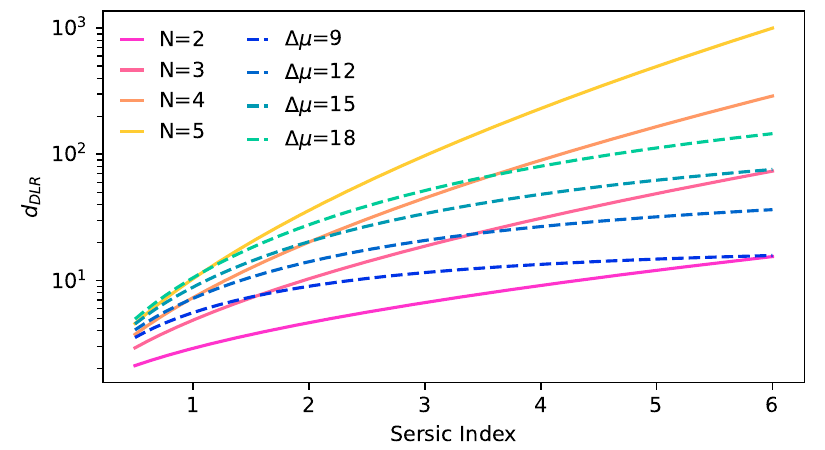}
\caption{Dependence of $\dDLR$ threshold on the S\'ersic index, under different probabilistic ($N\sigma$) and surface brightness ($\Delta\mu$) thresholds.
Both thresholds increase with the S\'ersic index, and the increase for the $N\sigma$ method is faster than for the $\Delta\mu$ method.
We select SNe where all galaxies within $1$ arcmin have $N\sigma>3$ or $\Delta\mu=12$ for visual inspection.
\label{fig:dlrthreshold}}
\end{figure}

In lieu of a classical, constant $\dDLR$ threshold, we propose two alternative techniques to identify possible hostless SNe: the constant fraction of enclosed light (or `$N\sigma$') method and the constant fraction of surface brightness (`$\Delta\mu$') method.

The $N\sigma$ method sets the $\dDLR$ threshold per the S\'ersic index ($\Nser$; \citealt{Sersic63}) so that the enclosed fraction of total flux inside the $\dDLR$ threshold is the same for galaxies of different light profiles.
We represent this enclosed light fraction using the number of standard deviations ($N$) under the empirical rule of statistics, such that the enclosed fraction equals the integrated probability within $(-N, N)$ for a normal distribution $\mathcal{N}(0, 1)$.
In analogue to hypothesis testing, assuming that a particular galaxy is the host galaxy, and that SNe are distributed following stellar light, then the enclosed light fraction at SN coordinates is complementary to the probability that a SN can happen at or beyond the radius (`$p$-value').
If the $p$-value is sufficiently small, or, correspondingly, the number of standard deviations ($N$) is large enough, then we can confidently reject the null hypothesis that this particular galaxy is the host. If all galaxies can be rejected in this way, then this SN is likely hostless.

Albeit statistically justified, this method has a clear drawback due to the dramatic variation of $\dDLR$ threshold with $\Nser$.
The same level of significance ($N$) implies a moderate $\dDLR$ cut for flatter light profiles (lower $\Nser$), but a large $\dDLR$ cut is required to achieve the same probability threshold for galaxies with steeper light profiles (higher $\Nser$).
As illustrated in Figure \ref{fig:dlrthreshold}, to achieve the same $3\sigma$ level of significance, the exponential profile ($\Nser=1$) requires a $\dDLR$ threshold of about $4$, while for a de Vaucouleurs’ profile ($\Nser=4$), the $\dDLR$ cutoff is close to $32$.
When excluding possible hosts, such a $3\sigma$ threshold is rigorous for galaxies with steeper light profiles but also tolerant to galaxies with flatter light profiles.

The second method (`$\Delta\mu$') that we propose is based on the assumption that the outskirts of a galaxy can be defined as the radius where the surface brightness drops to a constant fraction of the central region.
For the cutoff distance, we use the radius where the surface brightness, extrapolated from the best-fitting S\'ersic index, drops below a certain threshold compared to the mean surface brightness inside $R_\text{e}$.
This $\Delta\mu$ method is based on the local surface density and hence SN rate, so it is physically motivated. For a given S\'ersic index, we calculate the mean surface brightness within $R_\text{e}$ and the drop of surface brightness as a function of $\dDLR$.
Compared to the $N\sigma$ method, the $\dDLR$ cutoff of the $\Delta\mu$ method does not vary dramatically with the light profile;
meanwhile, the $\Delta\mu$ method compensates for the extended stellar light and SNe distribution beyond $R_\text{e}$ for high S\'ersic index galaxies.

Either method may have oversimplified the radial distribution of SNe in their host galaxies.
First, the radial light profiles of galaxies are usually more complicated than a single S\'ersic profile. There could be multiple structural components like stellar halos or extended envelopes. Such detailed morphological properties are, however, not available from the catalogues that we use here.
Second, SNe may not exactly follow the stellar light of their host galaxies, due to the radial gradient of stellar populations, and hence SN progenitors, in their host galaxies.
Assuming that SNe follow the light profile and using a single S\'ersic profile is therefore only an approximation of the SN radial distribution.

Nevertheless, we use both methods to select hostless SNe. We use $3\sigma$ as the probability threshold for the $N\sigma$ method and $12$ $\mathrm{mag\,asec^{-2}}$ for the $\Delta\mu$ method, respectively.
Most hostless SNe satisfy the criteria of both methods, while the sample selected only by the $N\sigma$ method contains more `Likely' cases per our visual inspection (Section \ref{sec:inspection}).

We use shape and size parameters from the LS catalogue to perform the $\dDLR$ cut.
However, similar parameters are not available for PS1 sources. The shape and size parameters of PS1 sources are measured using the moment of light, which differs from the profile-fitting photometry of the LS and contains seeing contributions.
We use the Kron radius \citep{Kron80} to derive similar $\dDLR$-like parameters for PS1 sources. To ensure comparable selection criteria across these two surveys, we construct simple k-Nearest Neighbour (kNN) regressors to estimate the cutting thresholds for PS1 sources.

We trained a series of kNN regressors using galaxies simultaneously observed by PS1 and LS, cross-matched with an angular distance limit of $2$ arcsec, inside $350$ circular fields (each with a $4$ arcmin diameter) randomly distributed in the overlap region of the LS and PS1.
We use the moment-based ellipticity, the Kron radius, and one of the colours in PS1 filters as the input; the desired output is the ratio of cutoff distance, calculated using LS source properties, to the Kron radius in the PS1 catalogue. The kNN regressors are constructed for $g-r$, $r-i$ and $i-z$ colours separately.
We then apply the trained kNN regressors to calculate the cutoff distance of galaxies in fields where only PS1 data are available.
We mainly use the kNN regressor trained using $g-r$ colour; in case if $g-r$ colour is corrupted or not measured for a galaxy, we also use regressors based on $r-i$ or $i-z$ colour.
In other words, to exclude possible unreported hosts outside the coverage of the LS, we use the average cutoff distances based on LS for PS1 sources with similar Kron radii, moment-based ellipticities, and optical colours.

There are other ways to associate SNe with possible host galaxies.
Notably, \citet{Gagliano21} use surface brightness gradient fields extracted from image cutouts to find host galaxies.
Also, \citet{Qin22} use machine learning-based ranking functions trained with known host-transient pairs to find hosts.
These methods are designed and optimized to find host galaxies when they present in images or catalogues, but they may not be effective to identify hostless SNe.

\renewcommand{\arraystretch}{1.1}
\begin{landscape}
\begin{table}
\caption{Table of Hostless Supernovae (The full table is available as online supplemental material)}
\label{tab:sourcetable}
\begin{tabular}{@{\extracolsep{6pt}}lcccccccccccc@{}}
\hline
Name & RA & Dec & $z$ & Type & Likely & Image & $g$   & $M_g$      & $M_g$     & $\log M_*$   & $\log M_*$     & $\log\text{SFR}$       \\ 
     &    &     &     &      &        &       &       & (Blue SED) & (Red SED) & (Passive)    & (Star-forming) &                        \\ 
     &    &     &     &      &        &       & (mag) & (mag)      & (mag)     & ($\mathrm{M}_\odot$)  & ($\mathrm{M}_\odot$)    & ($\mathrm{M}_\odot\,yr^{-1}$)   \\
\hline
2016aj  & 12:59:00.84  & -26:07:40.56 & 0.485 & SLSN-I &   & PS1 & 22.73 & -19.07 & -19.25 & 9.86 & 8.40 & -1.34 \\
2016ard & 14:10:44.558 & -10:09:35.42 & 0.203 & SLSN-I &   & PS1 & 23.13 & -16.80 & -16.66 & 9.34 & 8.01 & -1.84 \\
2016bfc & 16:41:09.71  & +50:58:20.9  & 0.050 & Ia     &   & PS1 & 23.07 & -13.61 & -13.35 & 7.93 & 6.69 & -3.22 \\
2016cwc & 16:45:34.37  & +22:54:30.43 & 0.130 & Ia     & * & PS1 & 23.13 & -15.74 & -15.53 & 8.28 & 7.12 & -2.76 \\
2016cwe & 16:45:27.45  & +37:04:49.79 & 0.150 & Ia     & * & PS1 & 22.73 & -16.41 & -16.04 & 7.56 & 6.26 & -3.61 \\
2016cwg & 15:52:36     & -08:36:12.14 & 0.100 & Ia     & * & PS1 & 22.98 & -15.77 & -15.15 & 8.39 & 7.04 & -2.84 \\
2016fmb & 21:35:25.48  & +21:19:12.47 & 0.070 & Ia     &   & LS  & 24.93 & -13.04 & -12.91 & 7.51 & 6.12 & -3.78 \\
2016iet & 12:32:33.23  & +27:07:15.49 & 0.068 & I      & * & PS1 & 23.01 & -14.31 & -13.89 & 7.98 & 6.72 & -3.18 \\
2017dah & 20:17:03.82  & -08:03:46.04 & 0.080 & Ia     &   & PS1 & 22.59 & -15.24 & -14.51 & 8.20 & 7.04 & -2.86 \\
2017dgk & 09:54:41.72  & -26:35:01.6  & 0.058 & Ic     &   & PS1 & 22.68 & -14.48 & -13.90 & 8.08 & 6.89 & -3.02 \\
\hline
\end{tabular}
\end{table}
\noindent \textit{Notes:} supernovae with asterisks in the \textit{Likely} column are likely hostless. Column $g$ is the extinction-corrected apparent magnitude limit ($5\sigma$) quoted using the survey indicated in \textit{Survey}. The $M_g$ columns show the $g$-band absolute magnitudes for two hypothetical SED models (red and blue), with joint constraints from other bands. The $\log M_{*}$ columns show the upper limits on stellar mass from SED fitting of the forced photometry, assuming two simple SED models (passive and star-forming). For the star-forming model, the implied upper limit on SFR is indicated in the last $\log\text{SFR}$ column.
\end{landscape}

\section{Catalog}
\label{appendix:sources}

The hostless SNe analyzed in this work is listed in Table \ref{tab:sourcetable}, along with the upper limits of host luminosity, stellar mass, and SFR, under different assumptions.

\section{Forced photometry using ZTF images}
\label{appendix:psfphotometry}

\begin{figure*}
\centering
\includegraphics[width=0.32\linewidth]{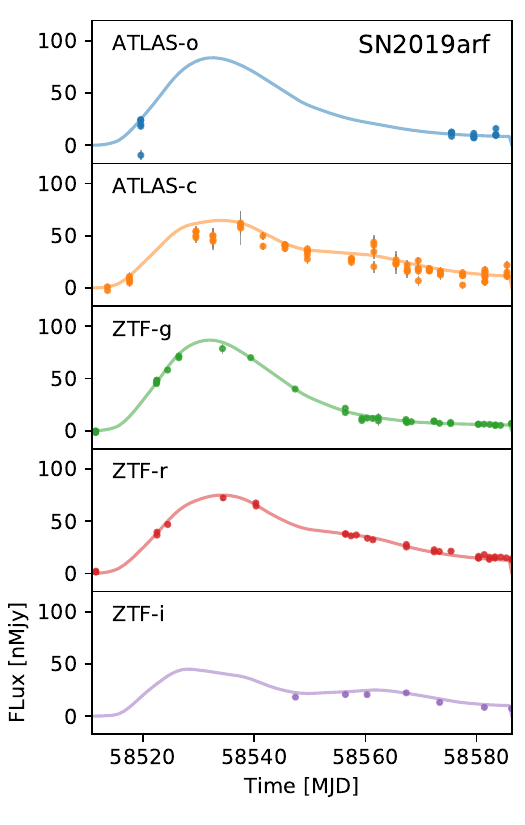}
\includegraphics[width=0.32\linewidth]{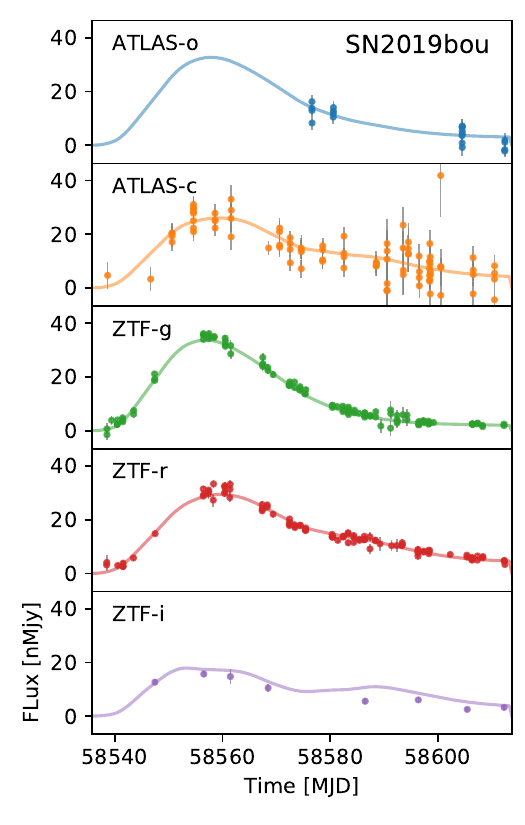}
\includegraphics[width=0.32\linewidth]{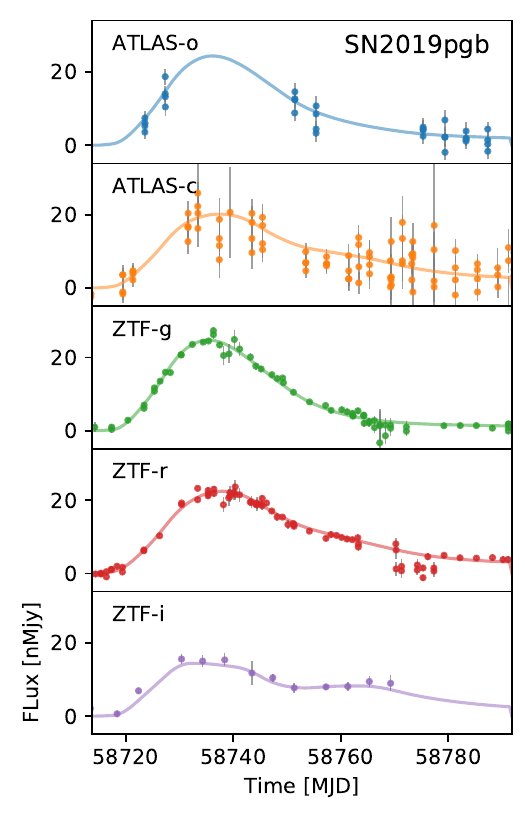}
\caption{
Examples of hostless SN Ia light curves.
The light curves in the ATLAS bands are constructed using their forced photometry service. In the ZTF bands, we calculate the flux using PSF photometry over difference images.
Solid lines show the best-fitting \textsc{SALT2} model \citep{Guy07, Guy10}, corrected for Galactic foreground extinction and the zero-point shifts due to any contamination of the reference images.
For clarity, we only show the phase range covered by the \textsc{SALT2} model.
Our light curve fitting effectively recovers the shapes of the light curves.
\label{fig:lightcurves}}
\end{figure*}

For ZTF photometry of hostless SNe Ia, the ZTF data releases\footnote{\url{https://www.ztf.caltech.edu/ztf-public-releases.html}}
contain light curves cross-matched from single-epoch PSF photometry catalogues obtained from the science exposures.
However, these light curves are not optimal for model fitting.
First, epochs of marginal detections and non-detections, which help constrain light curve parameters, are not included.
Second, PSF photometry on science images may suffer from contamination by the host galaxies, which could be an issue for our reference sample of hosted SNe.
Third, cross-matched single-epoch photometry may have an epoch-to-epoch variation in the source position, leading to potentially inconsistent photometry.

Therefore, instead of using the existing light curves, we carry out our own forced photometry using publicly-available image cutouts.
We obtain science, reference, and difference images around the SN coordinates, along with the effective PSF model of each epoch.\footnote{Data products are documented in detail at \url{https://irsa.ipac.caltech.edu/data/ZTF/docs/releases/}}
Epochs from 120 days before to 240 days after the date of discovery are included.
We fix the effective PSF model at the SN coordinates in the TNS and make the flux the only free parameter to fit.
We use the \textsc{DAOPHOT} algorithm \citep{Stetson87} implemented in \textsc{photutils} \citep{photutils} and the Levenberg-Marquardt solver to find the flux and its uncertainties.

We test whether the SN coordinates in the TNS are accurate enough for PSF photometry.
We align and stack difference images with positive detections ($\text{SNR}>3$) with inverse variance weighting (only including background RMS), while effective PSF models of the difference images are also stacked with identical per-epoch weights.
We then perform PSF photometry of the stacked images using the stacked PSF model, setting SN position as a free parameter.
The best-fitting SN coordinates are well in agreement with the given coordinates in the TNS, with typical offsets under $0.1$ arcsec.
ZTF images have a pixel scale of $1$ arcsec and typical FWHM of $2$ arcsec, and ATLAS images have a lower pixel scale of $1.86$ arcsec and FWHM of $3.8$ arcsec. The coordinates are therefore accurate enough for forced photometry.

To validate our forced PSF photometry, we also perform complementary aperture photometry using these images.
Using a fixed circular aperture of $7$ arcsec diameter, we measure the flux values at the SN positions and then correct the measured values using the enclosed fraction of flux estimated from the effective PSF model.
The corrected aperture flux is in agreement with our PSF photometry, showing no systematic shifts, although the flux uncertainties of PSF photometry are only about $70$ per cent of aperture photometry.
The difference of aperture and PSF photometry is attributable to the different methods used.
We notice that some reference images also contain the SNe flux due to the overlap of time periods between reference epochs and our SN sample.
This issue is offset by the flux baseline parameters, which are also free parameters in our fitting process.


\bsp	
\label{lastpage}
\end{document}